\documentclass[pra,preprintnumbers,nobalancelastpage,twocolumn,superscriptaddress,showpacs,nofootinbib]{revtex4-1}
\usepackage{amsmath, amssymb}
\usepackage{color}
\usepackage{graphicx}
\usepackage{subfigure}
\usepackage{booktabs}
\usepackage{ulem}
\usepackage[colorlinks=true, citecolor=blue, linkcolor=blue]{hyperref}

\newcommand{\ket}[1]{\ensuremath{\left | #1 \right \rangle}}
\newcommand{\bra}[1]{\ensuremath{\left \langle #1 \right |}}
\newcommand{\ketbra}[2]{\ensuremath{ \left | #1 \right \rangle \hspace{-0.1cm} \left \langle #2 \right | }}
\newcommand{\tr}{\text{tr}}
\newcommand{\lpo}[1]{\ensuremath{\hat \sigma^L_{#1} } }
\newcommand{\tracenorm}[1]{\ensuremath{{\| #1 \|}_1 } }
\newcommand{\average}[1]{\langle #1 \rangle}
\newcommand{\id}{\ensuremath{\hat {\boldsymbol 1}}}
\newcommand{\effdec}{\widetilde{\mathcal D}_t}
\newcommand{\nn}{\nonumber }

\newcommand{\mb}[1]{\ensuremath{\mathbf{#1}} }
\newcommand{\mc}[1]{\ensuremath{\mathcal{#1}} }
\newcommand{\bs}[1]{\ensuremath{\boldsymbol{#1}} }

\begin{document}
\def\bbm[#1]{\mbox{\boldmath$#1$}}

\title{Quantum memories with zero-energy Majorana modes and experimental constraints}

\author{Matteo Ippoliti}
\email{matteoi@princeton.edu}
\affiliation{Department of Physics, Princeton University, Princeton, NJ 08544, USA}
\affiliation{Scuola Normale Superiore, piazza dei Cavalieri 7,  I-56126 Pisa, Italy}

\author{Matteo Rizzi}
\affiliation{Universit\"at Mainz, Institut f\"ur Physik, Staudingerweg 7, D-55099 Mainz, Germany}

\author{Vittorio Giovannetti}
\affiliation{NEST, Scuola Normale Superiore and Istituto Nanoscienze-CNR, piazza dei Cavalieri 7, I-56126 Pisa, Italy}

\author{Leonardo Mazza}
\affiliation{D\'epartement de Physique, Ecole Normale Sup\'erieure / PSL Research University, CNRS, 24 Rue Lhomond, 75005 Paris, France}
\affiliation{NEST, Scuola Normale Superiore and Istituto Nanoscienze-CNR, piazza dei Cavalieri 7, I-56126 Pisa, Italy}

\begin{abstract}
In this work we address the problem of realizing a reliable quantum memory based on zero-energy Majorana modes in the presence of experimental constraints on the operations aimed at recovering the information.
In particular, we characterize the best recovery operation acting only on the zero-energy Majorana modes and the memory fidelity that can be therewith achieved. 
In order to understand the effect of such restriction, we discuss two examples of 
noise models acting on the topological system and compare the amount of information that can be recovered by accessing either the whole system, or the zero-modes only, with particular attention to the scaling with the size of the system and the energy gap.
We explicitly discuss the case of a thermal bosonic environment inducing a parity-preserving Markovian dynamics in which the memory fidelity achievable via a read-out of the zero modes decays exponentially in time, independent from system size.
We argue, however, that even in the presence of said experimental limitations, the Hamiltonian gap is still beneficial to the storage of information.
\end{abstract}


\maketitle

\section{Introduction \label{sec:intro}}

Notwithstanding the impressive progress in the experimental control of genuine quantum systems~\cite{Ladd_2005, Langer_2005, Balabas_2010, Tyryshkin_2012, Maurer_2012, Bar-Gill_2013, Pla_2013, Saeedi_2013, Muhonen_2014}, the reliable storage of quantum information over long times remains a problem of exceptional difficulty~\cite{Ladd_2010, Terhal_2015, Brown_2015}.
Among the ideas which have been put forward to solve this issue, proposals based on systems featuring topological order are among the most intriguing~\cite{Dennis_2002, Kitaev_2003, Nayak_2008}.
Central to these protocols is the encoding of one or more qubits in the degenerate ground states that appear in these models. Protection follows from the fact that such ground states are locally indistinguishable, and the information can be manipulated through some localized boundary modes, which may even feature non-Abelian statistics.

Due to their experimental relevance, topological superconductors featuring localized and unpaired Majorana modes have attracted considerable attention~\cite{Read_2000, Kitaev:2001}. 
Indeed, several reports accounting for the observation of experimental signatures of zero-energy Majorana modes have been recently published~\cite{Mourik_2012, Das_2012, Churchill_2013, Finck_2013, NadjPerge_2014}. 
At the same time, many proposals have discussed the possibility of observing them in cold atomic gases~\cite{Radzihovsky_2007, Sato_2009, DasSarma_2011, Jiang_2011, Diehl_2011, Kraus_2012, Nascimbene_2013, Kraus_2013, Buhler_2014}. 
The problem of assessing the robustness of quantum information protocols based upon these systems is now becoming of the highest relevance, and several works have already discussed many aspects of the problem~\cite{Goldstein_2011, Budich:2012, Bravyi_2012, Rainis:2012, Mazza:2013, Ho_2014, Ng_2013, Campbell_2015, Baranov_2014, DiVincenzo_2015_1, Baranov_2015, Carmele_2015, DiVincenzo_2015_2, Fulga_2013, Bravyi:2010}.

In this work we consider the following scenario: the experimentalist is able to create and cool a topological superconductor, but can only manipulate the degrees of freedom related to the Majorana fermions which appear in the system. 
This assumption is motivated by realistic implementation proposals for Majorana qubits and quantum memories \citep{Burrello_2013, Hassler_2011, Muller_2013, Alicea_2015}, 
and seems to be the mildest for discussing the near-future developments in the field. On one hand, experiments on many-body quantum systems have so far shown control over only a limited number of quantum degrees of freedom. On the other hand, if some quantum information has to be manipulated, it seems inevitable to assume some degree of control over the zero-energy Majorana modes; for the sake of simplicity, here we assume this control to be total. 
The upper bounds we derive apply {\it a fortiori} to more realistic situations of limited and imperfect control over the zero-modes.
The question we address is the following: how can one realize a reliable quantum memory in the presence of such constraints?

Assuming that the noise model is local, we explicitly characterize the best recovery operation~\cite{Mazza:2013} to be applied after the action of noise and determine the associated fidelity, both in general and under the aforementioned constraints.
In the latter case, which for clarity we dub \textit{decoding} rather than recovery,
we present an explicit recipe to be used in experiments in order to achieve the optimal fidelity.
It is worth stressing that in this work we do not consider possible protection operations to be applied during the action of noise, but rather consider noise as given and optimize the read-out.

In order to understand the effect of the experimental limitations, we discuss examples where an appropriate recovery operation acting over the whole topological system can retrieve the encoded information with $100\%$ fidelity  even if some form of noise has affected the system.
We show that a limited access to the Majorana modes of the system can compromise such possibility but we argue that in such cases information may still be protected by the presence of a significant energy gap in the topological Hamiltonian.

While the lack of protection of information in the constrained ``decoding'' case is in line with the general paradigm of quantum error correction, such behavior defies the expectationthat parity preservation should automatically protect information by making the Majorana modes immune to errors in the first place, which would make quantum error correction unnecessary. Our study highlights the fact that a control over the whole many-body system is indeed usually necessary.

Our study comprises two other additional results which we want to highlight here.
The first is about a topological system immersed in a thermal bosonic environment which induces a Markovian dynamics that does not change the parity of the number of fermions of the system. Notwithstanding this requirement, we find that the auto-correlation of a logical Pauli operator decays exponentially in time, with a time-scale given by the microscopic details of the system-environment coupling. 

The latter result comprises some exact relations for the case of a topological qubit encoded into two fully-decoupled Kitaev chains which are perturbed by a noise that conserves the parity of the number of fermions in each chain. In particular, we characterize the maximal amount of information which can still be recovered from the system after the action of the noise, thus generalizing the results in Ref.~\cite{Mazza:2013}. 

The work is organized as follows.
In Sec.~\ref{sec:enc} we present some introductory remarks about the encoding of a qubit in a topological system with zero-energy Majorana modes.
In Sec.~\ref{sec:eff-dyn} we discuss the action of the noise on the system and address the problem of its mathematical modeling, characterizing some of its general features. Using these results, we are able to characterize the best decoding operation and in particular its fidelity.
In Sec.~\ref{sec:ex:qq} we discuss the first example, namely the case of a quantum quench in the system. 
The second example is presented in Sec.~\ref{sec:bosonic}, where a master equation for a topological system in a bosonic environment is derived. In both cases, we explicitly compare the optimal recovery fidelity obtained by acting only on the zero modes (decoding) to that obtained by acting on the whole system (full recovery).
Finally in Sec.~\ref{sec:conclusions} we present our conclusions.

\section{Encoding a qubit in Majorana zero-modes \label{sec:enc}}

In this section we review some general aspects concerning the encoding of a qubit in Majorana zero-modes~\cite{Bravyi_2012, DiVincenzo_2015_2}.
We consider a generic topological superconductor with $N$ fermionic modes, without any assumption about the setup geometry.
We assume that the Hamiltonian is quadratic in the fermionic fields, and that the eigenmodes of the Hamiltonian include a group of four Majorana zero-modes $\{\hat m_i:\ i=1,2,3,4\}$ that satisfy the Clifford algebra $\{\hat m_i, \hat m_j \} = 2 \delta_{i,j}$ and are localized very far away from one another,
enough to make their overlap negligible for all practical purposes.
The remaining fermionic operators are assumed to be arranged in gapped eigenmodes represented by $N-2$ canonical annihilation (creation) Fermi operators $\{\hat b_i^{(\dagger)}: i = 1, \ldots, N-2\}$ which satisfy $\{\hat b_i, \hat b_j^{\dagger} \}=\delta_{i,j}$ and $\{ \hat b_i, \hat b_j\}=0$. Sometimes it will be convenient to use the related Majorana operators: $\hat \gamma_{i,1}  = \hat b_i + \hat b_i^\dagger$
and $\hat \gamma_{i,2} = -i(\hat b_i - \hat b_i^\dagger)$.
Analogously, the Majorana zero-modes can be algebraically combined to form non-local Dirac zero-modes $\hat g_0 \doteqdot \frac{1}{2}(\hat m_1+i\hat m_2)$, $\hat d_0 \doteqdot \frac{1}{2}(\hat m_3+i\hat m_4)$.
At this level, the pairing choice is entirely arbitrary.

A qubit can be stored in a well defined parity sector of the zero-energy two-fermion system, e.g. the two-dimensional space spanned by the logical vectors $\ket{0_L} = \ket{\Omega}$, $\ket{1_L} = \hat d_0^\dagger \hat g_0^\dagger \ket{\Omega}$. Here $\ket{\Omega}$ is the vacuum state of the fermionic system: $\hat b_i \ket{\Omega}=0$ for $i=1,\ldots,N-2$, $\hat g_0 \ket{\Omega }=\hat d_0 \ket{\Omega}=0$. 
We remark that four Majorana modes are necessary to store a non-trivial qubit:
with a single pair of Majorana modes (i.e. a single ``bi-localized'' Dirac mode) one can only encode a classical bit, as the fermionic parity superselection rule forbids coherent superpositions of the two degenerate ground states.

Denoting by $\hat \Pi_G$ the projector on the ground-space of the system, $\hat \Pi_G \doteqdot \prod_i \hat b_i \hat b_i^\dagger$, our prescription for the qubit encoding yields the following logical operators:
\begin{equation}
\left\{
\begin{aligned}
\lpo{0} & = \frac{1}{2} (\id -   \hat{m}_1  \hat{m}_2 \hat{m}_3 \hat{m}_4) \hat \Pi_G \;, \\
\lpo{1} & = \frac{i}{2}( \hat{m}_2  \hat{m}_3 +  \hat{m}_1  \hat{m}_4) \hat \Pi_G \;, \\
\lpo{2} & = -\frac{i}{2} (  \hat{m}_1  \hat{m}_3 -  \hat{m}_2 \hat{m}_4) \hat \Pi_G \;, \\
\lpo{3} & = -\frac{i}{2} ( \hat{m}_1 \hat{m}_2 +  \hat{m}_3  \hat{m}_4 )  \hat \Pi_G \;,
\label{eq:logical}
\end{aligned}
\right.
\end{equation}
which play the role of effective Pauli operators for our encoded qubit.
The $\sigma_j^L$s can be rewritten in a more compact way by introducing the fermionic parity operator $\hat P_f^G$ for the ground-space,
\begin{equation}
\hat P_f^G \doteqdot (-1)^{\hat g_0^\dagger \hat g_0 + \hat d_0^\dagger \hat d_0}
= -  \hat{m}_1  \hat{m}_2  \hat{m}_3  \hat{m}_4\;,
\label{eq:gs-parity}
\end{equation}
and letting $\hat \Pi_G^+ \doteqdot \frac{1}{2}(\id + \hat P_f^G) \hat \Pi_G$ denote the projector onto the even parity sector of the ground space.
Then, Eq.~\eqref{eq:logical} re-writes as 
\begin{equation}
\begin{aligned}
\lpo{0} & = \hat \Pi_G^+  \;, & 
\lpo{1} & = \hat \Pi_G^+ (i \hat m_2 \hat m_3) \;,  \\
\lpo{2} & = \hat \Pi_G^+ (- i \hat m_1 \hat m_3) \;, &
\lpo{3} & = \hat \Pi_G^+ (- i \hat m_1 \hat m_2)\;,
\end{aligned}
\label{eq:compact-notation}
\end{equation}
where $\hat \Pi_G^+$ commutes with the $i\hat m_j \hat m_k$ monomials. 
This formalism makes the Pauli algebra more evident.
It is worth stressing that the projector cannot generally be dropped in the qubit encoding operators $\{\lpo{i}\}$:
indeed, an initial qubit encoding which goes as $\hat \sigma_1^L \mapsto i \hat m_2 \hat m_3$ {\it et cetera} physically corresponds to a fully mixed (i.e., infinite-temperature) state  of the gapped modes, which also has mixed fermionic parity, and is a sub-optimal encoding choice.

From now on we assume that the experimentalist is able to create one such topological superconductor and to cool it to one of the two ground subspaces with fixed fermionic parity, e.g. $\hat \Pi_G^+$. We further assume that via control over the Majorana modes $\hat m_i$, it is possible to initialize the system in a desired pure state $\hat \rho (0) = \frac 12 \left( \hat \sigma_0^L+\vec v \cdot \vec{\hat{\sigma}}^L\right)$, where $|\vec v|=1$. Note that we can write $\hat \rho (0) = \hat \rho_0 \hat \Pi_G$, where $\hat \rho_0$ is a function of the zero-modes $\{ \hat m_i\}$ only.

As already mentioned, here and in the following we neglect the energy splitting which is associated to the overlap of the wavefunctions of the Majorana zero-modes \cite{Bravyi:2010}.
This energy splitting has the consequence that the localized Majorana modes are not exactly eigenmodes of the system and the logical operators in Eq.~\eqref{eq:logical} evolve in time.
However, since the Majorana wavefunction decays exponentially, the energy splitting scales exponentially in the distance between the zero modes, and it can be made arbitrarily small by increasing the system size. 
The time-scale associated to the evolution of the logical operators can be made arbitrarily long and in this article we consider the physical situation in which it can be safely neglected.

\section{Dynamics under noise and decoding of information \label{sec:eff-dyn}}

Once the qubit is initialized, the system is set free and noise and perturbations start acting on it. 
This drives the setup out of equilibrium by inducing a non-trivial and non-unitary dynamics, which may have disruptive effects on the encoded information. 
In order to describe  these disturbances, we introduce the decoherence channel $\mc D_t$, which returns the state of the global system at time $t$: $\hat \rho(t) = \mc D_t (\hat \rho(0))$.
The action of $\mc D_t$ includes the coherent evolution ruled by the topological Hamiltonian but does not entail any other additional passive protection mechanism.
For each value of $t$, $\mc D_t$ is a quantum channel, namely a completely-positive trace-preserving (CPTP) map on the states of the system~\cite{Nielsen, HOLEVOBOOK}.
This class of linear maps is known to describe the most general physical transformations of quantum states,
so that our formalism is completely general.
One additional physical requirement that we pose is that the dynamics must originate from a system-environment interaction that is local in space.

After a time $t$, the experimentalist attempts to ``decode'' the qubit, i.e. to recover the initially encoded information by reading only the four Majorana zero-modes. 
This limitation is physically relevant, since any protocol that allows the encoding in Eq.~\eqref{eq:logical} implies the ability to control the zero-modes; 
on the other hand, a generic global recovery operation may be significantly harder (or technically impossible) to carry out.

Since the {decoding} operation can only involve the zero-modes, the full decoherence channel $\mc D_t$
contains much more information than what is necessary for our purposes. 
As we shall wee below, we only need to know its restriction to the ground space, $\effdec$, which is obtained by tracing out the gapped modes $\{\hat b_i\}$:
\begin{equation}
\effdec (\hat \rho_0) \doteqdot \text{tr}_{\{\hat b_i\}} \left[ \mc D_t (\hat \rho(0 )) \right]=\text{tr}_{\{\hat b_i\}} \left[ \mc D_t (\hat \rho_0  \hat \Pi_G) \right].
\label{eq:effdec}
\end{equation}
The trace over the gapped modes of an operator $\hat O$ can be computed as follows: 
(i) expand $\hat O$ over the Hilbert-Schmidt orthogonal basis given by the monomials $\hat M_{\bs \alpha \bs \beta}$ in all the Majorana modes of the system, defined as follows:
\begin{equation}
\hat M_{\bs \alpha \bs \beta} \doteqdot 
\hat m_1^{\alpha_1} \cdots \hat m_4^{\alpha_4} \; 
\hat \gamma_{1,1}^{\beta_{1,1}} \;
\hat \gamma_{1,2}^{\beta_{1,2}} \cdots 
\hat \gamma_{N-2,1}^{\beta_{N-2,1}} \;
\hat \gamma_{N-2,2}^{\beta_{N-2,2}}\;,
\label{eq:monomials}
\end{equation}
where the $\alpha_i$ and $\beta_{i,j}$ coefficients can take values 0 or 1;
(ii) discard the monomials which contain at least one of the $\hat \gamma_{i,j}$;
(iii) multiply all the remaining monomials by $2^{N-2}$, which is the trace of the identity operator on the space of states of the gapped $\{ \hat b_i \}$ modes. 

Let us now motivate the introduction of the $\effdec$ channel. 
A general qubit recovery operation is a channel which maps  a state of the superconductor $\hat \rho$ to an abstract qubit state; it generally reads (see e.g.~\cite{Mazza:2013}):
\begin{equation}
 \mc R(\hat \rho) = \frac{1}{2} \left(\hat \sigma_0 +\sum_{i=1}^3 \text{tr} \hspace{-0.1cm} \left[\hat \rho \hat H_i\right] \hat \sigma_i\right)\;.
 \label{eq:recovery}
\end{equation}
The qubit is defined in terms of the abstract Pauli matrices $\hat \sigma_i$, $i=1,2,3$ and of the identity $\hat \sigma_0$, while the three coefficients which characterize the components of the Bloch vector are related to the measurement of three properly-selected (Hermitian) observables $\hat H_i$ (the request that $\mc R$ is a CPTP map imposes constraints on the $\{ \hat H_i\}$)
\citep{Mazza:2013, tesi:ippoliti}.
The physical constraint that the experimentalist can only manipulate the zero-energy modes $\{\hat m_i\}$ translates mathematically into the fact that the $\{ \hat H_i \}$ cannot be generic operators on the whole system; instead they must be polynomials in the zero-modes only.
One thus gets 
\begin{align}
\tr \hspace{-0.1cm} \left[\hat H_i \mc D_t(\hat \rho_0 \hat \Pi_G )\right] \; =& \; 
\tr_{\{\hat m_j\}} \hspace{-0.1cm} \left[ \hat H_i \tr_{\{\hat b_i\}} (\mc D_t(\hat \rho_0 \hat \Pi_G)) \right] \nonumber \\
\doteqdot& \; \tr_{\{\hat m_j\}} \hspace{-0.1cm} \left[ \hat H_i \effdec(\hat \rho_0) \right] \;,
\label{eq:trm}
\end{align}
so that $\effdec(\hat \rho_0)$ becomes the relevant object to study under this constraint. The goal of this Section is to characterize the three observables $\hat H_i$ which define the optimal operation $\mc R$ in Eq.~\eqref{eq:recovery} within the class of decodings.

\subsection{Constraints on the dynamics from locality and preservation of fermionic number parity \label{sec:constraints}}

In the Hamiltonian setting, the existence of Majorana zero-modes is linked to the fact that perturbations (i) are local and (ii) do not break the symmetry of the parity of the number of fermions~\cite{Hasan_2010, Qi_2011}.
It has often been suggested that similar constraints should also guarantee the robustness of information encoded into such modes~\cite{Akhmerov:2010, Hassler:2014}, although limitations have been pointed out \cite{Bravyi:2010}.
Leaving for later the discussion on the symmetry of the perturbation, let us here focus on the properties that a noise model should satisfy in order to be considered local.

In order to  introduce the concept of locality in our discussion, we shall consider decoherence channels which satisfy a Lieb-Robinson bound (LRB) for {all} pairs of fermionic (i.e. odd-degree monomials in the fermionic fields and combinations thereof) operators with support in distant regions $A$, $B$,
\begin{equation}
\{ \mc D_t^* (\hat O_A) , \hat O_B \} \simeq 0 \;,
\label{eq:flrb}
\end{equation}
and the {\it clustering property}
\begin{equation}
\mc D_t^* (\hat O_A \hat O_B) \simeq \mc D_t^* (\hat O_A) \cdot \mc D_t^* (\hat O_B)\;,
\label{eq:clustering}
\end{equation}
where $\mc D_t^*$ is the adjoint channel which describes time-evolution in the Heisenberg picture.
From now on, the approximation sign means that equality only holds up to LRB corrections, and thus in particular only for times $t \ll d_{AB}/v$, where $d_{AB}$ is the distance between regions $A$ and $B$ and $v$ is an effective group velocity.
Intuitively, Eq.~\eqref{eq:flrb} and \eqref{eq:clustering} imply that operators with distant, non-overlapping supports will take a long time to develop significant overlap and correlations.

The time interval within which the LRB corrections are negligible can be made arbitrarily long by increasing the size of the system and the distance between zero-modes $d_{AB}$. As a consequence, the approximation can be made arbitrarily accurate.
The following results on the dynamics of the Majorana modes will be derived within this framework and thus are valid only up to a finite time.
These points are quantitatively clarified in Appendix~\ref{app:local}.

Properties \eqref{eq:flrb} and \eqref{eq:clustering} can be used to prove that $\effdec$ is approximately diagonal in the basis given by the monomials $\{\hat M_{\bs \alpha} \doteqdot \prod_{j=1}^4 \hat m_j^{\alpha_j}, \, \bs \alpha \in \{0,1\}^4\}$.
In particular, we will show that:
\begin{subequations}
\label{eq:eff:deco}
\begin{align}
 \effdec(\hat m_i )\; \simeq& \; \lambda_i(t) \; \hat m_i; \qquad \qquad i = 1, 2,3,4; \label{eq:1}\\
 \effdec (\hat M_{\bs \alpha}) \; \simeq& \;\prod_{j=1}^4 \lambda_j^{\alpha_j}(t) \;\hat M_{\bs \alpha};
 \qquad \alpha \in \{0,1\}^4\label{eq:2}
\end{align}
\end{subequations}
up to LRB corrections, where $\lambda_j(t)$ are time-dependent scalar quantities. 
The result identifies the constraints on the dynamics of the ground states imposed by (i) the locality of the noise and (ii) the localization of the Majorana modes.
As an interesting corollary, the average parity in the ground-space is simply estimated by
\begin{equation}\label{eq:avparMR}
	\average{\hat P_f^G}_t = 
	\text{tr}[\hat P_f^G \hat \rho(t)] =
	\prod_{i=1}^4 \lambda_i(t).
\end{equation}
Moreover, the proof will return the proper definition of the adjoint channel:
\begin{equation}
\effdec^* (\hat O) = \tr_{\{\hat b_i\}} [\hat \Pi_G \mc D_t^*(\hat O)]
\label{eq:effdec-adjoint}
\end{equation}
[notice the different position of the projector $\hat \Pi_G$ in definitions \eqref{eq:effdec} and \eqref{eq:effdec-adjoint}].
The proof follows below; the reader uninterested in the technicalities can go directly to Sec.~\ref{sec:IIB}.

\textit{Proof of Eqs.~\eqref{eq:eff:deco} and~\eqref{eq:avparMR}.}
In order to present the demonstration, we shall introduce the adjoint (or Heisenberg-picture) reduced channel $\effdec^*$: 
if $\hat A$ and $\hat B$ are polynomials in the zero-modes,
then 
\begin{align}
\tr_{\{\hat m_i \}} [ \effdec(\hat A) \hat B ] 
& = \tr_{\{\hat m_i\}} [ \tr_{\{\hat b_i\}} (\mc D_t (\hat A \cdot \hat \Pi_G) ) \hat B] \nn \\
& = \tr[ \hat A \cdot \hat \Pi_G \mc D_t^* (\hat B)] \nn \\
& = \tr_{\{\hat m_i\}} [ \hat A \cdot \tr_{\{\hat b_i\}} (\hat \Pi_G \mc D_t^* (\hat B) )]\;,
\label{eq:proof}
\end{align}
which implies the definition in Eq.~\eqref{eq:effdec-adjoint}.

Let us now prove Eq.~\eqref{eq:1}:
the LRB \eqref{eq:flrb} implies that
$\mc D_t^*(\hat m_j) \simeq \hat m_j \cdot \hat Q_j(t)+ \hat Q'_j(t)$,
where $\hat Q_j(t)$ and $\hat Q'_j(t)$ are polynomials in the gapped modes supported (approximately) in the region of radius $vt$ centered around $\hat m_j$.
Notice that, since $\mc D_t$ is a physical evolution, $\mc D_t^*(\hat m_j)$ must be a fermionic operator, i.e. a linear combination of odd-degree monomials.
This forces $\hat Q_j(t)$ to be bosonic (i.e. even-degree) and $\hat Q_j^\prime(t)$ to be fermionic.
Therefore, 
\begin{align}
\effdec^*(\hat m_j) 
& = \tr_{\{\hat b_i\}} [\hat \Pi_G \mc D_t^*(\hat m_j)] \nonumber \\
&\simeq \hat m_j  \tr_{\{\hat b_i\}} [\hat \Pi_G  \hat Q_j(t)] +\tr_{\{\hat b_i\}} [\hat \Pi_G  \hat Q'_j(t)] \nn \\
& \doteqdot \lambda_j(t) \hat m_j\;;
\label{eq:lambda}
\end{align}
where we have used that $\tr_{\{\hat b_i\}} [\hat \Pi_G  \hat Q'_j(t)] = 0$ because $\hat \Pi_G \hat Q'_j(t)$ is a fermionic operator, and thus is traceless.

By the same reasoning, applying the clustering property~\eqref{eq:clustering}, 
one sees that 
$\effdec^* (\hat M_{\bs \alpha \mathbf 0}) = \Gamma_{\bs \alpha}(t) \hat M_{\bs \alpha \mathbf 0}$, where $\hat M_{\bs \alpha \mathbf 0}$ are monomials in the zero-energy modes only [see Eq.~\eqref{eq:monomials}].
The eigenvalue $\Gamma_{\bs \alpha}(t)$ is given by 
\begin{equation}
\Gamma_{\bs \alpha}(t) \doteqdot \tr_{\{\hat b_i\}} \big[\hat \Pi_G (\hat Q_1(t))^{\alpha_1} \cdots (\hat Q_4(t))^{\alpha_4} \big]\;.
\end{equation}
Let us recall that $\hat \Pi_G = \prod_i \hat b_i \hat b_i^\dagger$ is the ground-state projector: since the system (with the exception of the zero-modes) is gapped, correlations drop exponentially, and thus
\begin{equation*}
\Gamma_{\bs \alpha}(t)  = \langle  \hat Q_1(t)^{\alpha_1} \cdots \hat Q_4(t)^{\alpha_4} \rangle  
\simeq \langle  \hat Q_1(t)^{\alpha_1} \rangle \cdots \langle  \hat Q_4(t)^{\alpha_4} \rangle .
\end{equation*}
Recalling that $\tr_{\{\hat b_i\}} (\hat \Pi_G \hat Q_i(t)) = \langle \hat Q_i(t) \rangle \doteqdot \lambda_i(t)$, this yields
$\Gamma_{\bs \alpha} (t) = (\lambda_1(t))^{\alpha_1} \cdots (\lambda_4(t))^{\alpha_4}$ and proves that the reduced channel $\effdec^*$ inherits the clustering property \eqref{eq:clustering} from the full channel $\mc D_t^*$:
$\effdec^* (\prod_j \hat m_j) \simeq \prod_j \effdec^*(\hat m_j)$.

Finally, since the $\{\hat M_{\bs \alpha \mathbf 0}\}$ monomials form a basis of the space of operators on the zero-modes and are Hilbert-Schmidt orthogonal, 
we have that $\effdec^*$ is (up to LRB corrections) diagonal in an orthogonal basis, and thus self-adjoint. Therefore all the conclusions we derived for $\effdec^*$ apply equally to $\effdec$.

Using this result and the definition of the fermionic parity operator in Eq.~(\ref{eq:gs-parity}), it is now easy to prove the corollary \eqref{eq:avparMR}:
\begin{align}\label{eq:avparproof}
\average{\hat P_f^G}_t 
& = \text{tr}[ \hat \rho(t) \hat P_f^G ]
= \text{tr}[ \effdec ( \hat \rho(0) ) \hat P_f^G ] 
= \text{tr}[ \hat \rho(0) \effdec^* (\hat P_f^G) ] \nn \\
& \simeq  \prod_{i=1}^4 \lambda_i(t) \text{tr}[ \hat \rho(0) \hat P_f^G]
=  \prod_{i=1}^4 \lambda_i(t) \;.
\end{align}
This concludes the proof. $\blacksquare$

\subsection{Recovery fidelity and average parity \label{sec:IIB}}

After a time $t$ has elapsed, and noise and perturbations have acted on the system, the experimentalist attempts to recover the information previously encoded by applying a recovery operation~\eqref{eq:recovery} to the system. We define the fidelity of a recovery operation in the following way~\cite{Mazza:2013}:
\begin{equation}
 F[\mc R] = \int {\rm d}\mu_\phi \bra{\phi} \mc R \circ \mc D_t ( \, \ket{\phi}  \hspace{-0.1cm} \bra{\phi} \, ) \ket{\phi}
 \label{eq:recovery-fidelity}
\end{equation}
Here ${\rm d}\mu_\phi$ is the integration measure over the pure states of a qubit (i.e. the Bloch sphere).
By optimizing over the whole set of recovery operations $\mathfrak R$, we define the optimal recovery fidelity, an operational measure of the total amount of information left in the system~\cite{Mazza:2013}:
\begin{equation}
F_{\text{opt}} \doteqdot \max_{\mc R \in \mathfrak R} F[\mc R]\;.
\label{eq:global-maximization}
\end{equation}
In particular, given a decoherence channel, 
we are interested in the \textit{best} recovery fidelity that can be attained under the constraint that $\mc R$ only involves the manipulation of the Majorana zero-modes $\{\hat m_i\}$.
We thus limit the optimization to the set of ``decoding operations'' $\widetilde{\mathfrak R}$ that act on the zero-modes only (see the previous discussion), and define
\begin{equation}
 \widetilde{F}_{\rm opt} \doteqdot \max_{\mc R \in \widetilde{\mathfrak R} } F[\mc R]\;.
 \label{eq:maximization}
\end{equation}
Note that because of the maximization procedure in~\eqref{eq:maximization}, $\widetilde{F}_{\rm opt}$ provides a measure of the information which can be retrieved by manipulating the Majorana zero-modes only. More information is generally available by accessing the whole system ($F_{\text{opt}} \geq \widetilde{F}_{\text{opt}}$).

The use of $F_{\text{opt}}$ as a figure of merit is an important difference between our discussion and previous work in which a specific physically-motivated class of recovery operations is considered (e.g. stabilizer codes, minimal weight perfect matching)~\cite{Terhal_2015, Brown_2015}.
The optimal recovery fidelity represents a general upper limit to the performance of any possible recovery protocol, including all schemes previously considered in the literature. Thus, it might be used to test whether the known recovery operations are optimal or not,
and it is ideally suited to discuss general limitations of quantum memory models.
On the other hand, the recovery operation which achieves $F[\mc R] = F_{\text{opt}}$ does not generically have an interpretation in terms of known error-correction schemes. 
The introduction of $\widetilde{F}_{\text{opt}}$ is exactly motivated by the problem of giving a related figure of merit to which an operation can be associated.

Let us consider a scenario in which the $\lambda_j(t)$ from \eqref{eq:lambda} are all equal to $\lambda(t)$: in this simple symmetric case one can immediately compute the optimal decoding fidelity because $\effdec$ is effectively a simple qubit channel (see Appendix~\ref{app:single}):
\begin{equation}
\widetilde F_{\text{opt}} 
\simeq \frac{1+\lambda(t)^2 }{2}\;,
\label{eq:fidelity-parity:homo}
\end{equation}
which implies a relation between the decoding fidelity and the average parity in the ground-space:
$
\widetilde F_{\text{opt}} \simeq 
\frac{1}{2} \left( 1 +\sqrt{ \average{\hat P_f^G}_t} \right) 
$.

Moreover, at the end of this section we will show that this relation has a more general validity: if the $\lambda_j (t)$ are distinct, the following is true:
\begin{equation}
\widetilde F_{\text{opt}} 
\gtrsim
\frac{1}{2} \left( 1 +\sqrt{ \average{\hat P_f^G}_t} \right) \;,
\label{eq:fidelity-parity-lb}
\end{equation}
with the {approximate} equality sign only holding when $\lambda_j(t) = \lambda(t)$ $\forall\, j$.
To the best of our knowledge, this is the first mathematical relation linking the amount of information that can be stored in a system with zero-energy Majorana fermions and the parity of the fermionic number.

Note that the previous results for $\widetilde F_{\text{opt}} $ are true only up to LRB corrections. They are thus optimal only for times such that the effective light-cones originating from the Majorana zero-modes have not met yet. 
After such time, the value in Eq.~\eqref{eq:fidelity-parity:homo} is not guaranteed to be achievable. 

It is worth stressing that a decoherence channel $\mc D_t$ that preserves the fermionic parity of the whole system $\hat P_f = \hat P_f^G \cdot (-1)^{\sum_i \hat b_i^\dagger \hat b_i}$ does not necessarily preserve the ground-space parity $\hat P_f^G$. Even if no tunneling of individual fermions to or from the environment is allowed, transitions between the zero-modes and the gapped modes may take place.
Since in general the conservation of the fermionic parity of the system $\hat P_f$ is conjectured to be a necessary condition for the preservation of the information, Eq.~\eqref{eq:fidelity-parity-lb} points out that a limited control over the system might not be enough to benefit from it.

Nonetheless, since the Dirac modes $\hat b_i$ are gapped, one may argue that the transition of fermions from an excited mode $\hat b_i$ to e.g. $\hat g_0$, or vice-versa, should be strongly suppressed by the gap energy scale.
In parallel, within an open system setting, as long as the noise strength or environment temperature are small compared to the gap scale, one may also envision some form of protection. In Sec.~\ref{sec:ex:qq} and~\ref{sec:bosonic} we will discuss examples where these mechanisms are highlighted.

On the contrary, if no parity conservation is enforced on the noise model, then $\average{\hat P_f}_t \neq \average{\hat P_f}_{t=0}$ because e.g. of tunneling of fermions from the environment, and it is difficult to envision protection at any level~\cite{Goldstein_2011, Budich:2012, Rainis:2012, Mazza:2013, Ho_2014, Ng_2013, Campbell_2015, Baranov_2014, Baranov_2015, Carmele_2015}.

The proof of Eq.~\eqref{eq:fidelity-parity-lb} follows below; the reader uninterested in the technicalities can go directly to Sec.~\ref{subsec:best}.

\textit{Proof of Eq.~\eqref{eq:fidelity-parity-lb}.}
For simplicity, within this proof we use the notation $\lambda_j$ instead of $\lambda_j(t)$; the time-dependence is implicitly assumed.
We have, for any permutation $\{a,b,c,d\}$ of $\{1,2,3,4\}$, that
\begin{equation}
\effdec (i \hat m_a \hat m_b + i \hat m_c \hat m_d)  \simeq \lambda_a \lambda_b \, i \hat m_a \hat m_b + \lambda_c \lambda_d\, i \hat m_c \hat m_d\, .
\label{eq:abcd}
\end{equation}
Now, assuming without loss of generality that the $\lambda_j$ are such that $\lambda_1 \geq \dots \geq \lambda_4$, let us compute the recovery fidelity for the operation $\mc R \in \widetilde{\mathfrak R}$ defined in Eq.~\eqref{eq:recovery} with $\hat H$ matrices given by: 
\begin{subequations}
\label{eq:matrix:H:for:recovery}
\begin{align}
\hat H_1 & = \frac i2 (\hat m_2 \hat m_3\, + \hat m_1 \hat m_4), \\
\hat H_2 & = -\frac i2 (\hat m_1 \hat m_3- \hat m_2 \hat m_4)\,, \\ 
\hat H_3 & = - \frac i2 (\hat m_1 \hat m_2+ \hat m_3 \hat m_4)\,.
\end{align}
\label{eq:Hmatrices}
\end{subequations}
These matrices are related to the logical Pauli operators introduced in Eq.~\eqref{eq:logical} and describe an operation which simply attempts to decode the initially encoded information.
The associated recovery fidelity is 
\begin{align}
\widetilde {F}[\mc R] 
& = \frac{1}{2} + \frac{1}{12} \sum_{i=1}^3 \tr \big[\hat H_i \effdec(\lpo{i}) \big] \nonumber \\
& \simeq \frac{1}{2} \left[1+\frac{\lambda_1\lambda_2 + \lambda_3 \lambda_4 + \lambda_2 \lambda_3 + \lambda_1 \lambda_4 + \lambda_1 \lambda_3 + \lambda_2 \lambda_4 }{6} \right].\nonumber
\end{align}
The first line follows from the definitions \eqref{eq:recovery} and \eqref{eq:recovery-fidelity}, for a proof see e.g.~\citep{Mazza:2013}; the second line comes from applying \eqref{eq:abcd}.
The optimal fidelity will be at least as large as $\widetilde F[\mc R]$.
Now, invoking the inequality between arithmetic and geometric mean, one has
\begin{equation}
2 \widetilde F_{\text{opt}} [\effdec] 
\geq 1+ (\lambda_1^3 \lambda_2^3 \lambda_3^3 \lambda_4^3)^{1/6}
=  1+ (\lambda_1\lambda_2 \lambda_3 \lambda_4)^{1/2} 
\;.
\nonumber
\end{equation}
Recalling Eq.~(\ref{eq:avparMR}),
we get Eq.~\eqref{eq:fidelity-parity-lb}.
Moreover, if $\lambda_j = \lambda \ \forall j$, then the inequality is saturated and the choice of matrices in \eqref{eq:Hmatrices} is optimal.
$\blacksquare$

\subsection{Best decoding operation}\label{subsec:best}

The formalism which we developed also allows us to give a clear recipe for the best recovery operation to be performed by manipulating the Majorana zero-modes. Assuming for simplicity to be in the case $\lambda_j(t)=\lambda(t)$ discussed in Eq.~\eqref{eq:fidelity-parity:homo}, its operative definition is the following:
\begin{equation}
\mc R(\hat \rho) 
= \frac{1}{2} \left( \hat \sigma_0 + \sum_{i=1}^3 \text{tr} [\hat \rho \, \hat H_i] \hat \sigma_i \right)\;,
\label{eq:recovery:operation}
\end{equation}
where the $\{ \hat H_i \}$ are defined in Eq.~\eqref{eq:matrix:H:for:recovery} and the $\{ \hat \sigma_i \}$ are abstract Pauli matrices which describe the recovered qubit.

Eq.~\eqref{eq:recovery:operation} describes a simple decoding through operators related to the logical encoding in Eq.~\eqref{eq:logical}.
$\mc R$ thus requires the same experimental access to the Majorana modes which is initially used to perform the encoding.
More importantly, this puts on a sound ground the extent and the limitations of previous studies which have quantified information via looking at the fidelity of the logical qubits at different times~\cite{Goldstein_2011, Budich:2012, Rainis:2012, Baranov_2014, Baranov_2015, Carmele_2015}.

The generalization of Eq.~\eqref{eq:recovery:operation} to the case in which the  $\lambda_j(t) $ are distinct requires that rather than using the 
$\{ \hat H_i \}$ defined in Eq.~\eqref{eq:matrix:H:for:recovery}, one uses the monomial $i\hat m_a \hat m_b$ that yields the largest $\lambda_a \lambda_b$. 
For instance,
if at time $t$ we have $\lambda_1>\lambda_2>\lambda_3>\lambda_4$ and $\lambda_2 \lambda_3 > \lambda_1 \lambda_4$, then the optimal choice is $\hat H_1'=i \hat m_1 \hat m_2$, $\hat H_2'=i\hat m_2 \hat m_3$, $ \hat H_3'=i \hat m_3 \hat m_1$.

We remark that the operators $\hat H_i$ do not mutually commute, and hence are not compatible observables.
Therefore, the decoding operation \eqref{eq:recovery:operation} cannot be implemented as a straightforward sequence of three measurements, but rather the output qubit state has to be reconstructed by quantum state tomography \cite{Leonhardt_1995}, which requires a redundant encoding of the qubit.
Experimental schemes to measure the bi-localized operators $i \hat m_a \hat m_b$ have been proposed in various settings, including the top-transmon qubit \citep{Hassler_2011}, as well as the scalable Random Access Majorana Memory\cite{Burrello_2013}.

\section{Kitaev chain after a quantum quench}
\label{sec:ex:qq}

\begin{figure}[t]
\centering
\includegraphics[width = 0.9 \columnwidth]{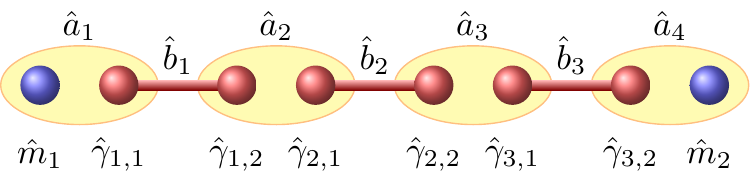}
\caption{(Color Online) Our notation for the Kitaev chain.
We set length $L=4$ for simplicity.
The large ellipses represent site modes $\hat a_i$;
the smaller spheres represent the Majorana modes.
Paired Majorana modes are labeled $\hat \gamma$, and the resulting bond eigenmodes (horizontal lines) are called $\hat b_i$.
Unpaired Majorana modes (at the edges) are labeled $\hat m$.
 \label{fig:modes}}
\end{figure}

Let us now study some explicit examples of decoherence channels $\mc D_t$, which 
in light of the previous discussion are chosen among those that preserve the parity of the number of fermions in the system. 
Our goals are 
(i) to compare the performance of recovery operations that act on the whole topological system to the performance of decodings of the zero modes, and 
(ii) to highlight the physical mechanisms which lead to protection of the information.
As a first simple example, in this Section we study the case of a perturbation represented by a quantum quench. The second example, which concerns a bosonic Markovian dynamics, is discussed in Sec.~\ref{sec:bosonic}.

We consider a pair of Kitaev chains in the topological phase~\citep{Kitaev:2001}.
Each chain is described by a Hamiltonian $\hat {\mathcal H}_{\rm K}$ and hosts two Majorana zero-modes at its edges; these modes can be combined as described in Sec.~\ref{sec:enc} to store a single qubit. The two chains are assumed to be identical and well separated, i.e. fully decoupled. 

The Hamiltonian of a single chain of length $L$ is:
\begin{equation}
 \hat {\mathcal H}_{\rm K} = \Delta \sum_{i=1}^{L-1} \left(- \hat a_{i+1}^\dagger \hat a_i^\dagger  + \hat a_{i+1}^\dagger \hat a_i + \textrm{H.c.} \right),
 \quad \Delta >0
 \label{eq:K:Ham:SP}
\end{equation}
where the $\{\hat a_i^{(\dagger)}: i=1, \ldots, L\}$ are canonical Fermi operators representing the annihilation (creation) of a fermion at the site $i$ of the chain. 
Hamiltonian~\eqref{eq:K:Ham:SP} is the Kitaev model with the parameters tuned to the so-called ``sweet point''~\citep{Kitaev:2001};
this peculiar choice represents an idealized situation which allows for simple calculations but being in the middle of a topological phase captures all the general features of the dynamics.
The Hamiltonian is diagonalized by a Bogolubov transformation, yielding $L-1$ eigenmodes with eigenenergy $\Delta$, denoted by $\{\hat b_i: i=1, \ldots, L-1\}$ and two unpaired Majorana edge modes
$\{\hat m_1, \hat m_2\}$:
\begin{equation}\label{eq:kitaev-ham}
 \hat{\mc H}_{\text{K}} = E_0+\Delta \sum_{i=1}^{L-1} \hat b_i^\dagger \hat b_i 
 = E_0' + \frac{\Delta}{2} \sum_{j=1}^{L-1} i \hat \gamma_{j,1} \hat \gamma_{j,2}\; .
\end{equation}
The $\hat b_i $ are the gapped modes of the system with energy $\Delta$ and are defined by $\hat b_i = \frac{1}{2} (\hat \gamma_{i,1} + i \hat \gamma_{i,2})$ in terms of the original Majorana modes (see Fig.~\ref{fig:modes} for a sketch).
$\Delta$ is the gap between the ground-space, spanned by $\ket{\Omega}$ and $\frac{1}{2}(\hat m_1 -i \hat m_2) \ket{\Omega}$, and the rest of the spectrum.

\begin{figure}[t]
 \includegraphics[width = \columnwidth]{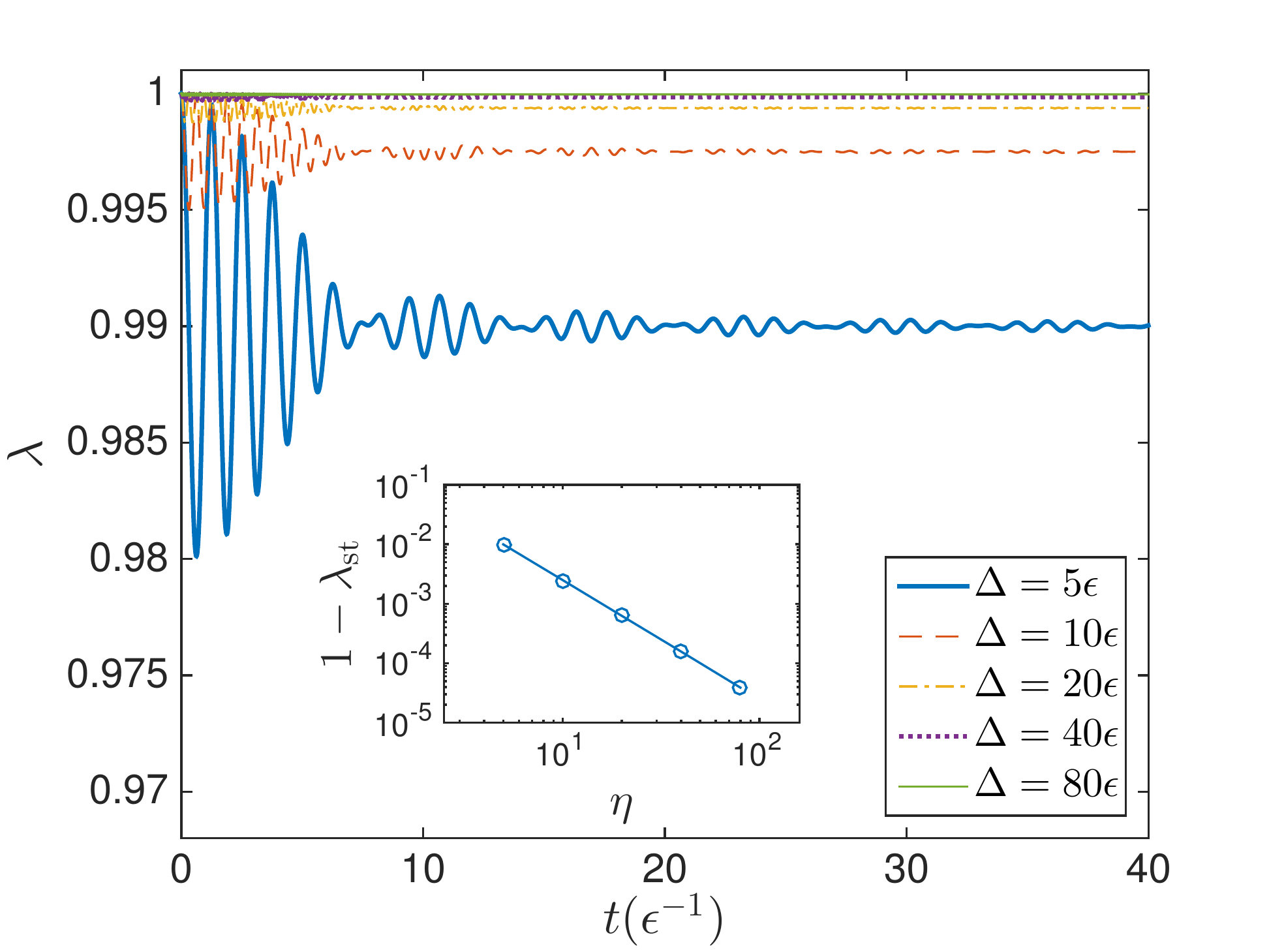}
 \caption{(Color Online) Evaluation of $\lambda(t)$ in Eq.~\eqref{eq:Quench:lambda} for the quantum-quench perturbation \eqref{eq:dec-unitary}, for $L = 100$, $\zeta = 1/2$ and several values of the gap of $\hat {\mathcal H}_{\rm K}$, $\Delta$.
The inset highlights that increasing $\Delta$ the steady value $\lambda_{\rm st}$ approaches $1$ as $(\epsilon/\Delta)^{2}$.
 }
 \label{fig:Lambda:Quench}
\end{figure}

At time $t=0$ the Hamiltonian is abruptly changed by adding a perturbation $\zeta \hat {\mathcal H}_{1}$: the system is thus set out of equilibrium and starts evolving according to a non-trivial unitary dynamics. 
Mathematically, the $\mathcal D_t$ channel is defined as:
\begin{equation}
\mathcal D_t \left( \hat \rho(0) \right) = 
 e^{-i (\hat {\mathcal H}_{\rm K} + \zeta \hat {\mathcal H}_1 )t}
 \hat \rho (0) 
 e^{i (\hat {\mathcal H}_{\rm K} + \zeta \hat {\mathcal H}_1 )t}.
 \label{eq:dec-unitary}
\end{equation}
For earlier work on this model, see Refs.~\cite{Bravyi_2012, Mazza:2013}.

We assume that $\hat {\mathcal H}_1$ does not couple the two chains: 
it thus suffices to focus on only one of them.
Such closed-system dynamics does not modify the parity of the number of fermions and moreover does not cause any information loss: the backward time evolution can in principle be used to restore the initial condition.
In order to engineer such operation, however, it is necessary to control the whole system; 
if the experimentalist has limited access to the system, a non-trivial fraction of the information may not be recoverable. 
Here we characterize this type of information loss for a {decoding} operation acting only on the zero-energy Majorana modes.

As a simple example of perturbation, we consider:
\begin{equation}
 \hat {\mathcal H}_1 = \epsilon \sum_{j=1}^L \hat a_j^\dagger \hat a_j;  \quad \epsilon >0.
\end{equation}
In this case, it is well-known that as long as $|\zeta |<2 \Delta/\epsilon$ the system remains topological, with zero-energy boundary Majorana modes. We will restrict ourselves to such a case.

In order to evaluate the optimal fidelity $\widetilde F_{\rm opt}$, we need to compute the $\lambda_j$ coefficients. 
Given the symmetry of the noise model, all the $\lambda_j$ coefficients are equal, therefore:
\begin{equation}
\label{eq:Quench:lambda}
\lambda_j(t) \equiv \lambda(t) = \frac 14 \text{tr} \left[ \mc D_t^* (\hat m_1) \hat m_1 \hat \Pi_G \right],
 \end{equation}
 where
 $
 \mc D_t^* (\hat m_1) = e^{i (\hat {\mathcal H}_{\rm K} + \zeta \hat {\mathcal H}_1 )t}
 \hat m_1 e^{-i (\hat {\mathcal H}_{\rm K} + \zeta \hat {\mathcal H}_1 )t}
$.
In Fig.~\ref{fig:Lambda:Quench} we plot the value of $\lambda(t)$ in Eq.~\eqref{eq:Quench:lambda}, computed by exploiting the properties of quadratic time evolutions, for several values of $\Delta>\epsilon$. 
We consider finite system sizes, $L \sim 100$, and finite times such that correlations arising from one edge have not yet reached the other one. $\lambda(t)$ then displays a clear steady behavior $\lambda_{\rm st}$, which we interpret as the value of the limit
$\lim_{t \to \infty} \lim_{L \to \infty} \lambda(t)$.
With the help of Eq.~\eqref{eq:fidelity-parity:homo} it is possible to relate the best decoding fidelity $\widetilde F_{\rm opt}$ to the value of $\lambda_{\rm st}$.
The fact that $\lambda_{\rm st}<1$ implies that the decoding is not perfect even in the limit $L \to \infty$. 
In presence of self-correction \cite{Preskill_2015}, the steady-state fidelity should scale as $1-e^{-c L}$.
Since in the regime considered here we find finite fidelity loss at infinite size, we can conclude that a simple decoding does not display self-correcting behavior.
However, the figure highlights the protecting effect of the gap $\Delta$: $|\lambda_{\rm st}-1| \sim (\epsilon/\Delta)^{2}$.

\begin{figure}[t]
 \includegraphics[width = \columnwidth]{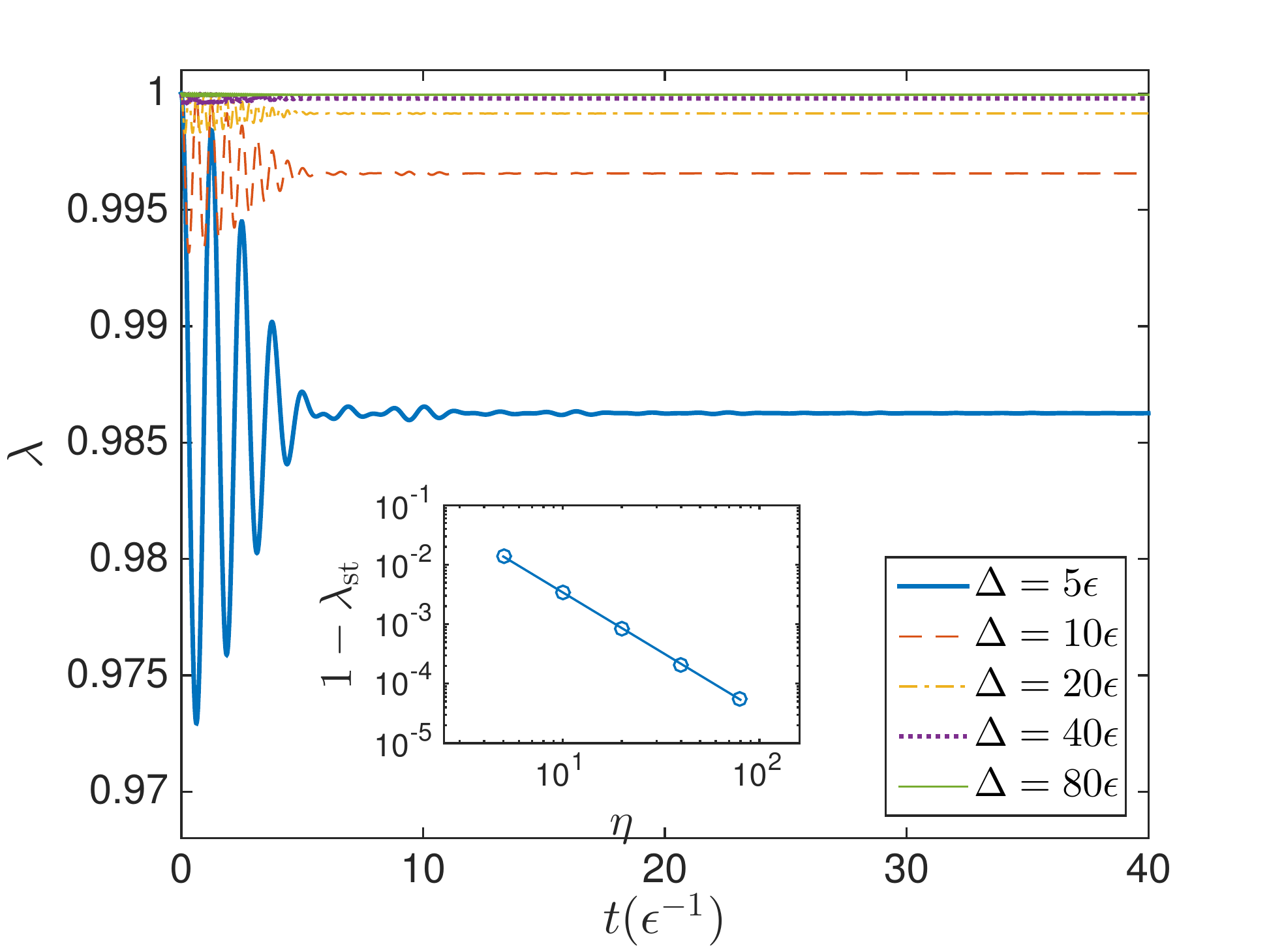}
 \caption{(Color Online) Evaluation of $\lambda(t)$ in Eq.~\eqref{eq:Quench:lambda} for the averaged quantum-quench perturbation in Eq.~\eqref{eq:pert:ave} for $L = 100$, and several values of the gap of $\hat {\mathcal H}_{\rm K}$, $\Delta$. The time evolution is generated by $\hat {\mathcal H}_{\rm K} + \zeta \hat {\mathcal H}_{1}$ and the average is performed over $50$ equispaced values of $\zeta$ between $0.02$ and $1$.
 The inset highlights that increasing $\Delta$ the steady value $\lambda_{\rm st}$ approaches $1$ as $(\epsilon/\Delta)^{2}$.
\label{fig:Lambda:Quench:Ave}
}
\end{figure}

Moving away from unitary dynamics, we can consider the case in which the parameter $\zeta$ is taken randomly from a distribution $g(\zeta)$ and the time evolution results from a classical average over such distribution:
\begin{equation}
 \mc D_t^* (\hat m_1) = \int g(\zeta) e^{i (\hat {\mathcal H}_{\rm K} + \zeta \hat {\mathcal H}_1 )t}
 \hat m_1 e^{-i (\hat {\mathcal H}_{\rm K} + \zeta \hat {\mathcal H}_1 )t} {\rm d} \zeta
 \label{eq:pert:ave}
\end{equation}
In Fig.~\ref{fig:Lambda:Quench:Ave} we show the results, which are qualitatively analogous to those obtained without averaging.
Note that the possibility of benefiting from full topological protection when acting on the whole system has been shown for this case in Ref.~\cite{Mazza:2013}.

Summarizing, the examples discussed in this Section show that a decoding of the zero-modes only does not enjoy any topological protection. The system is however stabilized by the gap of the topological Hamiltonian, with a protection that scales as a power law in the gap amplitude.

\section{Kitaev chain in thermal bosonic environment \label{sec:bosonic}}

We now consider a Markovian decoherence channel which arises from a thermal bosonic environment (i.e., an environment that conserves the parity of the number of fermions in the system). 
We begin deriving the master equation from first principles (see Refs.~\citep{Campbell_2015, tesi:ippoliti} for the case of fermionic environment).
Subsequently, we characterize the noise it induces in the system and discuss the optimal recovery fidelity, both with and without the experimental limitation of acting on the zero-modes only.

\subsection{Definition of the master equation}

The system considered in this example is analogous to that discussed in Sec.~\ref{sec:ex:qq}, namely it consists of two Kitaev wires which are separated and completely decoupled. Each Kitaev wire is surrounded by its own environment, so that even the noise dynamics cannot induce any correlation between the two chains. 
Similarly to the previous example, we assume that also the Hamiltonian does not couple the wires, so that it suffices to focus on the dynamics of a single chain.

\begin{figure}
\centering
\includegraphics[width=0.8\columnwidth]{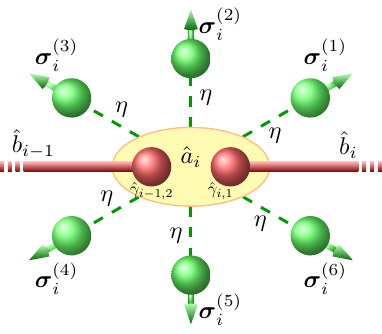}
\caption{(Color Online) Schematic of the system-environment coupling considered in section~\ref{sec:bosonic}, for a given site $i$ in the Kitaev chain (central ellipse, $\hat a_i$).
The arrows represent environment spins $\bs \sigma_i^{(1)},\dots, \bs \sigma_i^{({M})}$ (the case ${M}=6$ is shown; ideally ${M}\gg 1$).
Dashed lines represent the dephasing interactions $-\eta (\hat a_i^\dagger \hat a_i -\id/2) X_i^{(n)}$.
Thick lines represent pairing between Majorana modes.
 \label{fig:bose}}
\end{figure}

Let us first define the environment:
we assume that every site $i = 1,\dots,L$ of the chain is coupled to a large number of spins, ${M}$, via an interaction term which depends on whether the fermionic site is occupied. 
We thus introduce a collection of spin-1/2 $\hat {\bs \sigma}_i^{(n)} = (\hat X_{i,n},\hat Y_{i,n}, \hat Z_{i,n})$ and the Hamiltonian $\hat{\mc H}_{\text{env}} = \sum_{i=1}^L \sum_{n=1}^{M}  \omega_{i,n} \frac{1}{2} (I + \hat Z_{i,n} )$; 
the interaction with the chain is given by
\begin{equation}
\hat{\mc H}_{\text{int}} = -2\eta \sum_{i,n} (\hat a_i^\dagger \hat a_i -\id/2 ) \hat X_{i,n}\;.
\label{eq:bose-hint}
\end{equation}
While the coupling $\eta$ is homogeneous, the energy splitting of the spin levels, $\omega_{i,n}$, differs for each environment spin.
The $\{\omega_{i,n}\}$ energy gaps define an environment with density of states $f(\omega)$~\cite{Leggett_1987}. 
The situation is illustrated in Fig.~\ref{fig:bose}.

\begin{figure}
\centering
\includegraphics[width = 0.95\columnwidth]{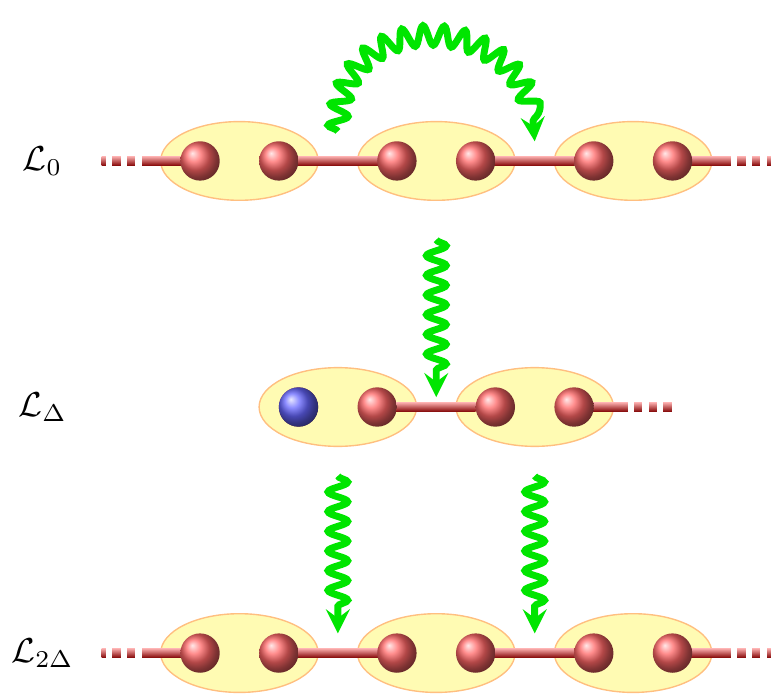}
\caption{(Color Online) Schematic of the three parts of the Lindbladian \eqref{eq:lindbladianDecomposition}.
(a) $\mc L_0$ moves an excitation from an occupied bond to a neighboring empty bond; no energy is exchanged between system and environment.
(b) $\mc L_\Delta$ creates (or destroys) an excitation in the first or last bond; a quantum of energy $\Delta$ (Hamiltonian gap) is transferred to the system from the environment (or vice versa).
(c) $\mc L_{2\Delta}$ creates (or destroys) a pair of excitations in two neighboring bonds; two quanta of energy $\Delta$ are transferred to the system from the environment (or vice versa).
\label{fig:lindbladians}
}
\end{figure}

In Appendix~\ref{app:mastereq2} we derive an effective master equation for this model in the weak-coupling limit $\eta \to 0$ and ${M} \to \infty$,  while ${M}\eta^2  $ is kept constant with value $g^2/\pi$. This, together with the assumption that there are no system-environment correlations in the initial state, ensures the Markovianity of the time evolution. 
For generality, the environment is assumed to be at inverse temperature $\beta$.
The associated Lindbladian defines the time evolution $\partial_t \hat \rho(t) = \mathcal L[\hat \rho(t)]$ 
and the decoherence channel $\mathcal D_t [\hat \rho(0)] = e^{t \mathcal L} [\hat \rho(0)]$.
It consists of three parts:
\begin{equation}
\mc L = \mc L_0 + \mc L_\Delta + \mc L_{2\Delta} \;, 
\label{eq:lindbladianDecomposition}
\end{equation}
which are described here below, and also schematically represented in Fig.~\ref{fig:lindbladians}.
The situation is analogous to that discussed in Refs.~\cite{DiVincenzo_2015_1, DiVincenzo_2015_2} for the Kitaev chain and in several works for the toric code coupled to a thermal environment~\cite{Terhal_2015, Brown_2015}.

The term
$\mc L_0$ is given by processes that do not exchange energy between the system and the environment. 
Those include the closed-system Hamiltonian dynamics $\mc H$ (which, as we argue in Appendix~\ref{app:mastereq2}, generally gets a ``Lamb-shift'' correction, $\hat{\mc H}_{LS}$, due to the environment), and the incoherent propagation of an existing excitation from an occupied bond to a neighboring empty bond, $\hat L^{(0)}$:
	\begin{subequations}
	\begin{align}
	\hat{\mc H} & = \Delta \sum_{i=1}^{L-1} \hat b_i^\dagger \hat b_i\, + \hat{\mc H}_{LS} \; ; \\
	\hat L_i^{(0)} 
	&   =  \sqrt{2\phi(0)} (\hat b_i^\dagger \hat b_{i-1} + \hat b_{i-1}^\dagger \hat b_i )\;.
	\label{eq:Lind0MR}
	\end{align}
	\end{subequations}
The coefficient 
\begin{equation}
 \phi(\omega) \doteqdot \frac 12 g^2 f(|\omega|) \big[1+\tanh(\beta\omega/2) \big]
\end{equation}
is introduced in Appendix~\ref{app:mastereq2} in the derivation of the master equation, and represents an effective coupling that takes into account the environment density of states and thermal occupation number. Note that it satisfies the detailed-balance equation $\phi(\omega)/\phi(-\omega) = e^{\beta \omega}$~\cite{Leggett_1987}.
The Hamiltonian correction $\hat{\mc H}_{LS}$ is given in Eq.~\eqref{eq:LShamiltonian} and does not gap the zero-modes.

The term $\mc L_\Delta$ is given by dissipative processes that require the exchange of a unit $\Delta$ of energy from the system to the environment, or \textit{vice versa}. These are:
\begin{align}
\hat L_1^{(\Delta)} & = \sqrt{ \phi(\Delta) } \, i\hat m_1 \hat b_1\;, &
\hat L_1^{(-\Delta)} & = \sqrt{ \phi(- \Delta) } \, i\hat m_1 \hat b_1^\dagger \;, \nn \\
\hat L_2^{(\Delta)} & = \sqrt{ \phi(\Delta) } \, i\hat m_2 \hat b_{L-1} \;, &
\hat L_2^{(-\Delta)} & = \sqrt{ \phi(- \Delta) } \, i\hat m_2 \hat b_{L-1}^\dagger \;.
\end{align}
These operators describe the {incoherent} {creation or annihilation of an excitation} next to an edge, and involve explicitly the Majorana zero-modes.

Finally, the term $\mc L_{2\Delta}$ represents dissipative processes that require the exchange of {two} units of energy $\Delta$ between the system and the environment.
These processes are represented by Lindblad operators
	 \begin{align}
	 \hat L_{i}^{(2\Delta)} 
	 & = \sqrt{\phi(2\Delta)}\, \hat b_{i-1} \hat b_i\;, & 
	 \hat L_{i}^{(-2\Delta)} 
	 &  =  \sqrt{\phi(-2\Delta)}\, \hat b_{i-1}^\dagger \hat b_i^\dagger \;. \nonumber
	\end{align}
These describe the {incoherent} {creation of a pair of excitations} in neighboring empty bonds, or the annihilation of a pair of excitations in neighboring occupied bonds. Note that the $\hat L_i^{(0)}$ and the $\hat L_i^{(2\Delta)}$ do not depend on the zero-energy modes.

As a final consistency check, we can easily see that all the Lindblad operators we derived are quadratic in the Majorana modes (and thus bosonic). As we prove in Appendix~\ref{app:pp}, this is equivalent to having a master equation that preserves fermionic number parity.

As anticipated earlier in the work, the gap of the Kitaev Hamiltonian can protect the information if it is large compared to temperature.
If we set $\Delta \gg 1/\beta$, the processes that pump energy  from the environment into the system are suppressed exponentially in $\beta \Delta$. Additionally, processes that drain energy from the system or that do not exchange energy act trivially because there are no excitations to move or annihilate in the ground space. In Sec.~\ref{sec:finiteT} we consider a super-Ohmic density of states $f(\omega) \propto \omega^2$~\cite{Leggett_1987} and, resorting to numerical techniques, we discuss the temperature-dependence of the typical memory time-scale.

Before doing so, we analyze a model which allows for an interesting preliminary analysis obtained setting $\beta = 0$ and $f(2\Delta) \ll f(\Delta) \ll f(0)$.
This latter assumption implies the relation $\phi(0) \gg \phi(-\Delta) \gg \phi(-2\Delta)$, which mimics the low-temperature regime, in which high-energy states become unavailable.
However, setting $\beta=0$, the asymmetry between $\phi(\Delta)$ and $\phi(-\Delta)$ characteristic of finite temperatures is not introduced, and thus  no qualitative analogy should be in principle expected. We discuss this case because, thanks to several algebraic simplifications, it is possible to obtain analytical results. 
Similarly to the full thermal case, we obtain that in the limit $f(\Delta), f(2\Delta) \to 0$ the information is fully recoverable, as expected.
In the following subsections we study the recoverability of the information when acting on the zero-modes or on the whole system for both cases.

\subsection{Analytical model: $\beta=0$ and $f(2\Delta) \ll f(\Delta) \ll f(0)$}

\subsubsection{Decoding fidelity}

Let us discuss the properties of the channel $\effdec$ related to the Markovian dynamics
$e^{t(\mc L_0 + \mc L_\Delta + \mc L_{2\Delta} )}$ just introduced, for $\beta = 0$. 
In particular, we are interested in the behavior of the optimal decoding fidelity in the $L \gg 1$ limit.
Starting from the formula in Eq.~\eqref{eq:fidelity-parity:homo}, we
compute the $\lambda_1(t)$ coefficient:
\begin{align}
\lambda_1(t) 
=& \frac{1}{2} \tr_{\{\hat m_i\}} (\hat m_1 \effdec (\hat m_1 )) 
 = \frac{1}{2} \tr \left( \hat m_1 \mc D_t [\hat m_1 \hat \Pi_G ] \right) = \nonumber \\
 = & \frac{1}{2}  \tr \left( \mc D_t^{*} [\hat m_1] \hat m_1 \hat \Pi_G  \right)
 = \frac{1}{2} \tr \left( e^{t \mc L^* }[\hat m_1] \, \hat m_1 \hat \Pi_G \right)\;.
\label{eq:lambda2}
\end{align}
Since the channel considered here does not couple the two chains, we can focus on a single wire.
In particular, Eq.~\eqref{eq:lambda2} refers to a single chain and the normalizing factor of 2 is such that $\frac{1}{2} \tr \hat \Pi_G = 1$.

Since $\beta = 0$, we have
$\phi(\omega) = \phi(-\omega)$, and thus $\hat{L}^{(\omega)\dagger} = \hat{L}^{(-\omega)}$ for all Lindblad operators.
Therefore $\mc L^*$ is equal to $\mc L$, except for the sign of the Hamiltonian term (the only anti-Hermitian term in $\mc L$).
As $\hat m_1$ commutes with $\hat{\mc H}$, $\mc L^*$ acts on $\hat m_1$ just like $\mc L$:
\begin{equation}
\mc L^* [\hat m_1] 
 = \hat m_1 \mc L_0 [\id] +\mc L_\Delta [\hat m_1] + \hat m_1 \mc L_{2\Delta} [\id] 
 = -g^2 f(\Delta) \hat m_1\;,
\label{eq:ME:terms}
\end{equation}
Thus $ e^{t\mc L^*} [\hat m_1] = e^{-g^2 f(\Delta) t} \hat m_1$, and
\begin{equation}
\lambda_1(t)  = \frac{1}{2} \tr( e^{-g^2 f(\Delta) t} \hat m_1\cdot \hat m_1 \hat \Pi_G )
= e^{-g^2 f(\Delta) t}\;.
\label{eq:lambda-analytical}
\end{equation}
It is easy to see that the exact same argument applies to $\hat m_2$, and thus to both zero-modes of the other chain, so that all four $\lambda_i(t)$ coefficients are the same.
The optimal decoding fidelity~\eqref{eq:fidelity-parity:homo} is therefore:
\begin{equation}
 \widetilde{F}_{\text{opt}}(t) = \frac{1+e^{-2g^2 f(\Delta) t}}{2}.
 \label{eq:F:opt.ME}
\end{equation}

A number of remarks are in order.
First, let us stress that the formula for $\widetilde{F}_{\rm opt}(t)$ in Eq.~\eqref{eq:F:opt.ME} does not depend on the size of the system, $L$,
as it is derived assuming the $L \to \infty$ limit. 
This implies the absence of self-correcting behavior, which would give $\widetilde{F}_{\rm opt}(t) \to 1$ as system size is increased.

Second, in the limit $f(\Delta) \to 0$, we observe perfect recoverability. Since $F_{\rm opt} \geq \widetilde F_{\rm opt}$, in this limit no error occurs in the system because, independently from the complicated dynamics which may arise in the bulk, the zero-energy modes are unperturbed.
The corruption of information is thus due to the Lindbladian term $\mathcal L_\Delta$.

\subsubsection{Comparison with full recovery operations}

Let us now compare the optimal decoding fidelity~\eqref{eq:F:opt.ME} with the optimal recovery fidelity achievable by acting on the whole system.
In Appendix~\ref{app:best:best:best} we prove that in the presence of a decoherence channel which (i) does not couple the two Kitaev chains and (ii) preserves the fermionic parity of each chain, the optimal recovery fidelity is:
\begin{equation}
 F_{\text{opt} }(t) =  \frac{2}{3} + \frac{1}{12} \left(\| \mc D_t(\hat m_1 \hat \Pi_G ) \|_1 \right)^2\;.
 \label{eq:optfidSingleChain}
\end{equation}
The formula is a simple generalization of the optimal recovery fidelity provided in Ref.~\cite{Mazza:2013} for the case of two Kitaev chains with noise acting on only one of them.
Interestingly, the fact that the quantum channel $\mathcal D_t$ acts in a product way on the two chains yields a significant advantage, as for the computation of $F_{\rm opt}(t)$ one only needs to consider a single chain. From a numerical point of view, this is crucial for obtaining some meaningful size scaling in the cases in which the methods of fermionic linear optics cannot be applied (such as this one).

\begin{figure}[t]
\includegraphics[width=\columnwidth]{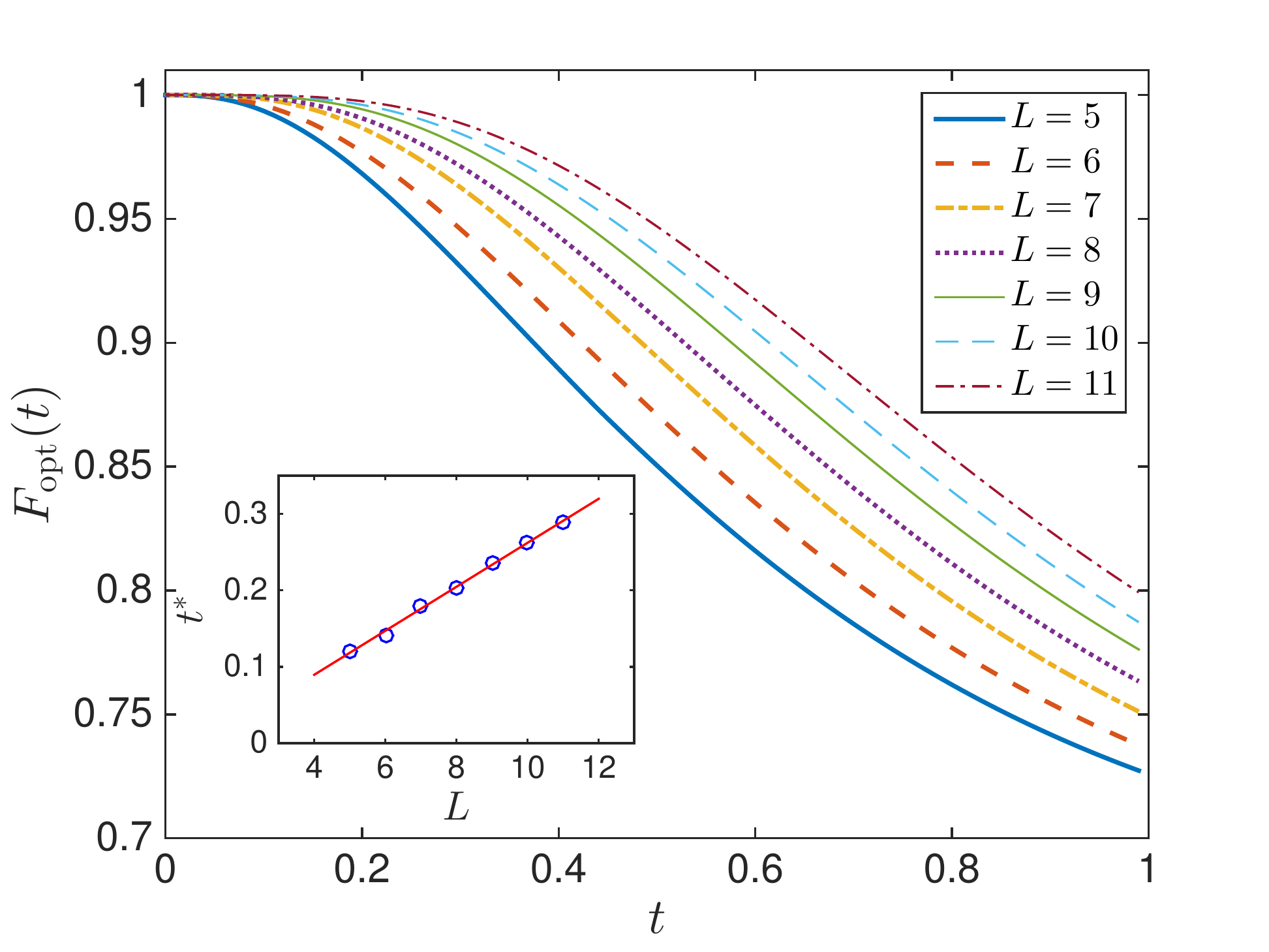}
\caption{(Color online)
Optimal recovery fidelity for a system of two Kitaev wires, for several values of the wire length $L$.
We set $\Delta = 1$ as the unit of energy (and inverse time), $g^2 f(0) = 4$, $f(\Delta) = 0.3 f(0)$, $f(2\Delta) = 0.3^2 f(0)$.
The inset shows the linear scaling of coherence time $t^*$ with the chain length $L$.
We define $t^*$ as the time at which $F_{\text{opt}}(t^*) = 0.99$.
}
\label{fig:ME1:Data}
\end{figure}

We numerically compute the optimal recovery fidelity with a Runge-Kutta (RK) integration of the differential equation $\partial_t \hat  A = \mathcal L [\hat A]$. The exponential scaling of the dimension of the Hilbert space limits our computation to sizes of order $L=11$ (length of a single chain). Results are reported in Fig.~\ref{fig:ME1:Data}, where it is shown that the value of $F_{\rm opt}(t)$ depends on the size.

In order to quantify how the memory properties of the system depend on the size $L$, we fix an arbitrary threshold value $F_{\rm thr}<1$ and define the memory time $t^*(L)$ as the first time at which $F_{\rm opt}(t)$ crosses such value. In the inset in Fig.~\ref{fig:ME1:Data} we take $F_{\rm thr} = 0.99$ and demonstrate that the scaling of $t^* (L)$ is linear in $L$. 
The increase of coherence time with system size represents a form of protection, even though $t^*(L)$ does not show the exponential scaling with $L$ which is considered a signature of self-correcting behavior.

The obtained linear behavior $t^*(L) \sim L$ is rather surprising because at the single particle level the master equation induces a diffusive dynamics for the excitations, which yields a scaling $t^*(L) \sim L^2$. 
In order to understand the discrepancy between the expected diffusive behavior and the seemingly ballistic behavior we observe, we have studied the problem in some significant limits.
The quadratic (diffusive) scaling has been recovered by restricting the dynamics to the subspace of states with energy smaller than $2 \Delta$ (and thus no more than two $\hat b_i$ excitations in the chain) [not shown]. 
Additionally, setting the rate of creation and annihilation of excitation pairs to zero, 
$\phi(2\Delta)=\phi(-2\Delta) = 0$,
the full many-body calculation shows a quadratic scaling $t^*(L) \sim L^2$ [not shown]. 
Both cases agree with the intuitive picture of fidelity being lost when an excitation created by $\mc L_\Delta$ at one edge is diffused through the bulk by $\mc L_0$ until it can annihilate with an excitation coming from the opposite edge.
This shows that the transition from diffusive to ballistic regime
(i) is due to the possibility of creating pairs of excitations in the bulk,
and (ii) it is a many-body problem which cannot be recast in terms of
simple unbiased random walks.
We leave the development of a simple microscopic model giving rise to this unexpected ballistic behavior for future work.

\subsubsection{Comparison with other figures of merit}

Before concluding the study of this setup, we consider two  figures of merit which are defined independently from recovery operations and have already been discussed in the relevant literature.

The first figure of merit is related to the temporal resilience of the Majorana fermions, and was employed for instance in Ref.~\cite{Carmele_2015}:
\begin{equation*}
 \theta_1 (t) = \text{tr} [i \hat m_1 \hat m_2 \, \hat \rho(t)]
\end{equation*} 
Here, $\hat \rho (t) = e^{t \mathcal L} [(1+ i \hat m_1 \hat m_2 )\hat \Pi_G/2]$.
Recalling Eq.~\eqref{eq:trm} we obtain:
\begin{equation*}
 \text{tr} \left[ i \hat m_1 \hat m_2 \mathcal D_t (\hat \rho_0 \hat \Pi_G) \right] = \text{tr}_{\{ \hat m_j \}} \left[ i \hat m_1 \hat m_2 \effdec (i \hat m_1 \hat m_2 / 2) \right] .
\end{equation*}
We thus obtain:
\begin{equation*}
 \theta_1(t) = \lambda(t)^2 = e^{- 2 g^2 f(\Delta) t}.
\end{equation*}

As a second figure of merit, we consider the autocorrelation time of the Majorana fermions, which was considered for example in Ref.~\cite{Goldstein_2011}:
\begin{equation*}
 \theta_2 (t) = \left|\text{tr} [i \hat m_1 \hat m_2 \, \mc D_t^* (i \hat m_1 \hat m_2) \hat \rho_0] \right|
 \end{equation*}
  Here the trace is over two chains and the initial state is taken to be $\ket{0_L}$: $\hat \rho(0) = {\frac{1}{2}(\lpo{0} +\lpo{3})} = \frac{1}{4}(\id -i \hat m_1 \hat m_2 - i \hat m_3 \hat m_4 - \hat m_1 \hat m_2 \hat m_3 \hat m_4 ) \hat \Pi_G^a \hat \Pi_G^b$.
  Chain $b$ can be traced out, yielding
  \begin{equation*}
   \theta_2 (t) = \left| \frac{1}{2} \text{tr}_a [(\id - i \hat m_1 \hat m_2) \, i \hat m_1 \hat m_2 \, \mc D_t^* (i \hat m_1 \hat m_2) \hat \Pi_G^a ] \right|
  \end{equation*}
Finally, the calculation of $\mc D^*_t (\hat m_1 \hat m_2)$ simply follows from Eq.~\eqref{eq:ME:terms}:
\begin{equation*}
\theta_2 = \left| \frac{1}{2} \text{tr}_a [\hat m_1 \hat m_2 \, e^{-2g^2 f(\Delta) t} \hat m_1\hat m_2 \hat \Pi_G ] \right| = e^{-2 g^2 f(\Delta) t}.
\end{equation*}

The similarity between these results and the constrained optimal fidelity $\widetilde F_{\rm opt}$~\eqref{eq:F:opt.ME} underpins with information-theory techniques the methods which have been previously proposed.
Additionally, this shows that even a bosonic environment can represent a danger for the quantum information encoded in Majorana zero-modes if control over the system is experimentally limited.

\subsection{Environment at finite temperature \label{sec:finiteT}}

\begin{figure}[t]
 \includegraphics[width = \columnwidth]{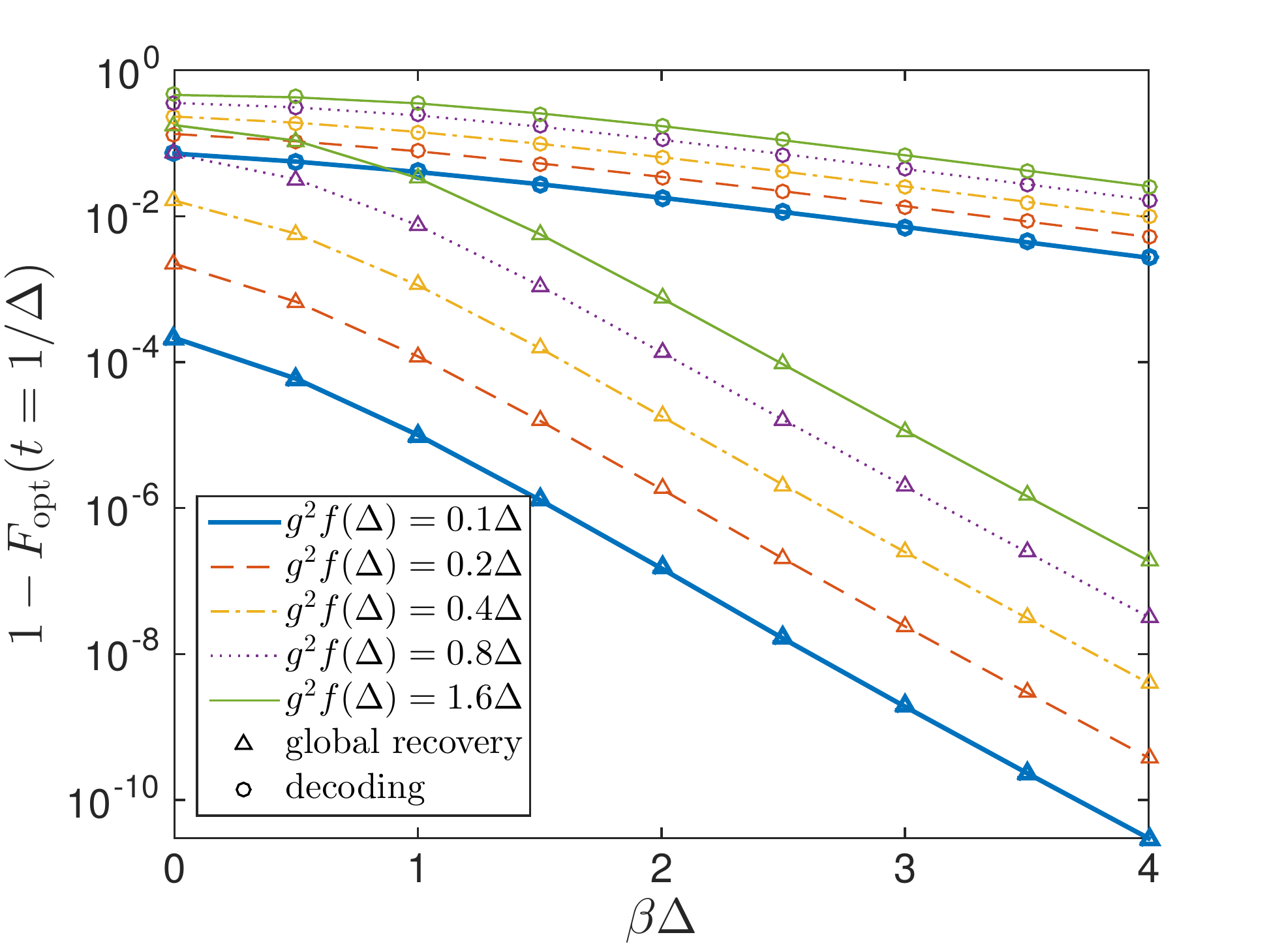}
 \caption{(Color Online) Fidelity loss at $t=1/\Delta$ as a function of inverse temperature $\beta$, for several values of the noise parameter $g^2 f(\Delta)$.
$\Delta$ is kept constant. 
We assume $f(\omega) \propto \omega^2 e^{-\omega/\Omega}$ with the cutoff $\Omega \doteqdot 5\Delta$, so that $f(2\Delta) = 4 e^{-1/5} f(\Delta) \simeq 3.3 f(\Delta)$ and $f(0) = 0$.
We show both the case of a globally optimized recovery operation (triangles) and that of a decoding of the zero-modes (circles).
Though in both cases the fidelity loss is exponentially suppressed in $\beta $, the performance of the unconstrained recovery operation is significantly better than that of the constrained one:
the fidelity loss drops as $\sim e^{-1.8 \beta \Delta}$ in the former case, and as $e^{-0.4 \beta \Delta}$ in the latter.
}
 \label{fig:finiteT}
\end{figure}

Let us now turn to the study of the temperature effects on the memory efficiency.
As an example, 
we consider $f(\omega) \propto \omega^2 e^{-\omega/\Omega}$, which describes a super-Ohmic bath~\cite{Leggett_1987}, where $\Omega$ is a high-energy cutoff necessary to make the integrals~\eqref{eq:integrals:for:LS} in the ``Lamb shift'' Hamiltonian~\eqref{eq:LShamiltonian} convergent.
For consistency, we work in the regime $\Omega \gg \Delta$ and $\Omega \gg 1/\beta$.

We then assume $\beta > 0$, so that there is an asymmetry between processes that pump energy into the system and those that drain it out of the system. 
Because of this asymmetry, $\hat m_1$ is no longer eigenmode of the Lindbladian $\mc L^*$, and thus the analytical argument that we used to prove Eq.~\eqref{eq:F:opt.ME} breaks down.
Thus, both fidelities $F_{\rm opt}$ and $\widetilde F_{\rm opt}$ must be computed numerically.

We thus consider a pair of chains of length $L = 8$ and we compute both kinds of optimal recovery fidelity at time $t=1/\Delta$.
Given the chosen functional form for $f(\omega)$, the relevant noise parameters obey $f(0) = 0$ and $f(2\Delta) = 4 e^{-\Delta/\Omega} f(\Delta)$. Note that in Refs~\cite{DiVincenzo_2015_1, DiVincenzo_2015_2} a slightly different thermal environment is considered.
We then vary independently $\beta$ and the noise intensity $g^2 f(\Delta)$, while $\Delta$ is kept constant, and plot the resulting fidelity loss $1-F_{\text{opt}}(t=1/\Delta)$.

We remark that the exact vanishing of $f(0)$, while rather unrealistic and fine-tuned, is a useful limiting case to study: 
since it completely inhibits one of the diffusion mechanisms in the chains (the other being $\mc L_{2\Delta}$), it should improve the memory performance with respect to the $f(0)\neq 0$ case.
Thus, any negative results obtained in this scenario will hold {\it a fortiori} for generic case.

The results are represented in Fig.~\ref{fig:finiteT}. 
As expected, a low environment temperature ensures the protection of information: 
as $T \to 0$ (i.e. $\beta \to \infty$), there is no energy in the environment nor in the initial state of the chain, so the system state is confined to the zero-energy sector, where no local interaction couples $\hat m_1$ and $\hat m_2$; thus information survives indefinitely.
This ideal regime is approached exponentially in $1/T$, what is usually called an Arrhenius activation law~\cite{Brown_2015}; however, we observe that the characteristic energy scale is different depending on whether one has full access to the system or is confined to the zero-modes. 
Once again, most of the information loss observed when acting on the edge modes only would actually be recoverable via a global operation, and only a tiny fraction is fatally lost to the environment. 
The ratio between the information lost by looking only at the edge modes and that really unrecoverable is expected to increase exponentially in $\beta\Delta$.

Another significant result that emerges from Fig.~\ref{fig:finiteT} is that at a fixed time the fidelity loss scales only polynomially with the noise intensity $g^2 f(\Delta)$, while it is suppressed exponentially in $\beta \Delta$. Note the difference with Eq.~\eqref{eq:F:opt.ME}, where at fixed time the scaling with $g^2 f(\Delta)$ is exponential (no comparison with the temperature scaling is possible because there we set $\beta=0$).
This means that increasing $\Delta$, the gap of the Hamiltonian, leads to an exponential suppression of the thermal fluctuations, and to a polynomial increase in the noise ($g^2 f(\Delta) \sim \Delta^2$): the result is an Arrhenius scaling typical of Hamiltonian protection.

As we already mentioned, in the finite-temperature case the analytical arguments we used for the $\beta  = 0$ case break down, so we cannot obtain a simple generalization to \eqref{eq:lambda-analytical} for $\lambda(t)$.
Nonetheless, it is possible to evaluate $\lambda(t)$ perturbatively for short times.
Going back to the definition \eqref{eq:lambda2} and expanding to first order in $t$ yields
\begin{equation}
\lambda(t) = 1 + \frac{t}{2} \tr \big[ \mc L^* (\hat m_1) \hat m_1 \hat \Pi_G \big] + O(t^2)\;;
\end{equation}
it is easy to verify that 
\begin{equation}
\mc L^*(\hat m_1) = -2 g^2 f(\Delta) \hat m_1 \frac{\id + \tanh(\beta\Delta/2) i \hat \gamma_{1,1} \hat \gamma_{1,2}}{2}\;,
\end{equation}
so that 
\begin{equation}
\lambda(t) = 1- g^2 f(\Delta) \left(1-\tanh \left(\frac{\beta\Delta}{2}\right)\right) t + O(t^2)\;.
\end{equation}

Let us stress that this results returns the correct short-time expansion of~\eqref{eq:F:opt.ME} for $\beta = 0 $.
In the $\beta \Delta \gg 1$ limit, it yields:
\begin{equation*}
 \lambda(t) \simeq 1 - 2 g^2 f(\Delta) e^{-\beta \Delta} t
\end{equation*}
Thus, as long as $f(\Delta)$ is polynomial in $\Delta$~\cite{Leggett_1987}, 
setting $\Delta \gg k_B T$, i.e. increasing the Hamiltonian gap, leads to an Arrhenius-type protection of information in both the recovery and decoding scenarios, with different activation energies. 
This is a size-independent effect that shows, again, the absence of self-correction.
This conclusion would only be reinforced by removing the constraint $f(0)\neq 0$ from the bath spectral density.

\section{Conclusions \label{sec:conclusions}}

In this article we have discussed the realization of a reliable quantum memory based on a topological superconductor hosting unpaired zero-energy Majorana modes. Motivated by the current technical limitations in the manipulation and control of such quantum fluids, we have addressed the problem of understanding whether a reliable storage of quantum information in such systems is possible with minimal assumptions regarding read-out capabilities.
In particular, we have characterized the best recovery operation to be applied after the action of the noise when the control over the system is strictly limited to the zero-energy Majorana modes where the information is initially encoded. 
We have identified the fidelity of such operation, quantifying the information which remains in the initial encoding degrees of freedom, and provided experimental recipes for the optimal decoding.

Explicit calculations for two examples of noise models are also presented. 
The examples (a quantum quench in the Kitaev Hamiltonian and a bosonic thermal environment) are chosen because in those cases it is possible to prove that a recovery operation acting on the whole system is substantially more effective than a simple decoding of the Majorana zero-modes only. 
Thus their analysis best highlights the handicap introduced by the minimal experimental requirements.

While in both examples a decoding of the zero-modes does not enjoy any topological protection (i.e. finite fidelity loss persists even in the ideal infinite-size limit), we have highlighted that it benefits from the presence of the Hamiltonian energy gap, displaying Arrhenius-type behavior. This is true even when the environment has a density of states which increases with the energy (we have explicitly considered a super-Ohmic behavior).

More generally, our results rely on the existence of a LRB in local unitary and open Markovian dynamics. 
Although in this article we have not entered the paradigm of non-Markovian dynamics, the employed formalism might still be used for such situations. 
It is rather intriguing to speculate that a LRB might exist also in those cases, as even non-Markovian dynamics should originate from a unitary dynamics in the larger system-environment setup: 
if such underlying unitary dynamics is local, it must obey the rule of light-cone propagation of correlations.
Thus, we expect our findings to extend to more general scenarios.

Finally, in order to bridge our article to the previous literature, we have compared our results to figure of merits different from the best recovery fidelity. Remarkably, the open-system dynamics studied in this work provides a non-trivial example of \textit{bosonic} environment where all the considered figures of merits display an exponential decay over time without size protection. 
This feature is thus not due to any fine-tuning in the figures of merit, and raises interesting questions about the role of parity in protecting the information stored in the system. 
Our study stresses that parity-conservation is not sufficient to protect information if the system cannot be properly manipulated.
The precise conditions under which the encoding into Majorana fermions can protect information from a parity-preserving noise require further investigations.

\acknowledgments

We gratefully acknowledge feedbacks and comments by M.~Burrello and R.~Fazio.
We are personally indebted to C.-E.~Bardyn for pointing out to us the diffusive nature of excitations under thermal noise.
L.~M. and M.~R. acknowledge invaluable discussions with J.~I.~Cirac and M.~D.~Lukin in the early stages of the work.
L.M. acknowledges financial support from Italian MIUR through FIRB project RBFR12NLNA.
L. M. was supported by LabEX ENS-ICFP: ANR-10-LABX-0010/ANR-10-IDEX-0001-02 PSL*.
 
\appendix

\section{Local noise models \label{app:local}}

In this Appendix we discuss some properties of local Markovian noise models, which are used throughout the paper.
We start from the original Lieb-Robinson bound (LRB) \citep{Lieb:1972}, an inequality which bounds the propagation of correlations in closed spin systems with {\it local} interactions:
\begin{equation}
\| [\hat O_A(t), \hat O_B(0) ] \|_{op} \leq cV \| \hat O_A \|_{op} \, \| \hat O_B \|_{op} e^{ - ( d_{AB} -vt)/\xi} \;.
\label{eq:lrb-spin}
\end{equation}
Here $\hat O_A$, $\hat O_B$ are operators supported in distant regions $A$ and $B$, respectively; 
$d_{AB}$ is the distance between such regions;
$V$ is the volume of the largest such region;
$c$, $v$ and $\xi$ are model-dependent constants.
This means that there is an effective light cone, with ``speed of light'' $v$, that describes the propagation of correlations from a local region $A$. Correlations are not strictly zero outside the light cone, but rather fall off on a distance scale $\xi$.

Since Eq.~\eqref{eq:lrb-spin} deals with closed systems, the time-evolved operator (in the Heisenberg picture) is $\hat O_A(t) = \hat U^\dagger (t) \hat O_A \hat U (t)$. 
However, it was shown by Poulin \citep{Poulin:2010} that the same result holds if the unitary evolution is replaced by a Markovian channel (again in the Heisenberg picture) $\hat O_A(t) = \mc D_t^* (\hat O_A)$, provided the Markovian dynamics is {\it local}.
Locality in this context means that the Lindbladian $\mc L$ can be decomposed as $\sum_{X \subset \Lambda} \mc L_X$, where $X$ are subregions of the system with diameter at most $d$ and each $\mc L_X$ has Lindblad operators supported in $X$.

We observe that the same reasoning that leads to the LRB \eqref{eq:lrb-spin} for spin systems can be mapped exactly to a fermionic setting \citep{Hastings:2004}, provided the Markovian dynamics is parity-preserving:
by the result of Appendix~\ref{app:pp}, parity preservation implies bosonic Lindblad operators, so that the commutation between distant operators (the key ingredient for the proof in the spin case) is recovered.
One can thus prove that
\begin{align}
[ \mc D_t^* (\hat O_A^b), \hat O_B^b] & \simeq 0 \;, &
\{ \mc D_t^* (\hat O_A^f) , \hat O_B^f \} & \simeq 0\;,
\label{eq:lrb-fermions}
\end{align}
where $\hat O_A^b$, $\hat O_B^b$ are distant {\it bosonic} operators, $\hat O_A^f$, $\hat O_B^f$ are distant {\it fermionic} operators, and the approximation symbol is meant as a shorthand for the right hand side of the inequality \eqref{eq:lrb-spin}, i.e. it holds up to LRB corrections.

Finally, one last useful property of local, Markovian, parity-preserving noise models that we use is the following:
\begin{equation}
\mc D_t^* (\hat O_A \hat O_B) \simeq \mc D_t^* (\hat O_A) \mc D_t^*(\hat O_B)\;,
\label{eq:app-clustering}
\end{equation}
where the approximation again means ``up to LRB corrections''.
We call \eqref{eq:app-clustering} a {\it clustering property}, as it implies the clustering of correlations for distant operators on uncorrelated (product) initial states:
\begin{equation}
\average{\hat O_A \hat O_B }_t \simeq \average{\hat O_A}_t \average{\hat O_B}_t\;,
\label{eq:clustering-correlations}
\end{equation}
where $\average{\hat O}_t \doteqdot \tr [\hat O \mc D_t(\hat \rho_A \hat \rho_B) ] = \tr[\mc D_t^* (\hat O) \hat \rho_A \hat \rho_B] $, 
for all states $\hat \rho_A$ ($\hat \rho_B$) supported in region A (B). 
Property \eqref{eq:clustering-correlations} was proven by Poulin \citep{Poulin:2010} for all local Markovian channels of {\it spin} systems;
however, the same arguments are straighforwardly mapped to fermion systems, provided the dynamics is parity-preserving (i.e. all Lindblad operators are bosonic -- see Appendix~\ref{app:pp}), as that ensures the commutation between parts of the Lindbladian supported in distant regions, which is the key argument used in the proof.
In \citep{Poulin:2010}, Eq.~\eqref{eq:app-clustering} is actually proven as an intermediate step towards Eq.~\eqref{eq:clustering-correlations}.

\section{Optimal single-qubit recovery operation \label{app:single}}

Here we prove a formula for the optimal recovery fidelity of a single qubit exposed to noise.
It is known that a general single qubit noise can be expressed as follows \citep{Ruskai:2002}:
\begin{equation}
\frac{\id +  \mb n \cdot \bs \sigma }{2} \xrightarrow{\mc D} \frac{\id +  (\Lambda \mb n + \mb b) \cdot \bs \sigma }{2}\,,
\end{equation}
where $\Lambda$ is a matrix with singular values $\upsilon_1$, $\upsilon_2$, $\upsilon_3 \in [0,1]$, namely:
$\Lambda = R \cdot D \cdot S$, with $D = \text{diag}(\upsilon_1,\upsilon_2,\upsilon_3)$ and $R,\ S \in O(3)$.
The $S$ rotation can be eliminated by a suitable unitary redefinition of the Pauli matrices:
$\bs \sigma \mapsto S \bs \sigma$, $\mb n \mapsto S\mb n$, $\mb b \mapsto S\mb b$ preserves $\mb n \cdot \bs \sigma$ and $\mb b \cdot \bs \sigma$, while it yields
\begin{equation}
(\Lambda \mb n) \cdot \bs \sigma = \mb n^T S^T D R^T \bs \sigma 
\mapsto \mb n^T D R^T S^T \bs \sigma\;.
\end{equation}
We can then simply call $SR \doteqdot R$ and effectively get rid of $S$.

The optimal recovery operation is now the unitary gate that undoes the last rotation $R$: 
\begin{equation}
\frac{\id +  \mb n \cdot \bs \sigma }{2} \xrightarrow{\mc R} \frac{\id +  (R^T \mb n ) \cdot \bs \sigma }{2}\,,
\label{eq:qubit:recovery}
\end{equation}
which gives an average recovery fidelity
\begin{align}
F[\mc R] 
& = \int {\rm d}^2 \mb n\ \text{tr} \left( \frac{\id +   \mb n \cdot \bs \sigma }{2} \frac{\id +  R^T(\Lambda \mb n + \mb b) \cdot \bs \sigma }{2} \right) = \nn \\
& = \frac{1}{2} \int {\rm d}^2 \mb n\ [1+\mb n^T (R^T\Lambda \mb n + R^T\mb b)]= \nn \\
& = \frac{1}{2} \left[1+ \int {\rm d}^2 \mb n \ \mb n^T \text{diag}(\bs \upsilon) \mb n \right] =\nn \\
& = \frac{1}{2} \left( 1 + \frac{\upsilon_1+\upsilon_2+\upsilon_3}{3} \right)\;.
\end{align}

As discussed in Ref.~\cite{Mazza:2013}, the fidelity of a recovery operation must obey the following inequality:
\begin{equation}
F_{\text{opt}} \leq \frac{1}{2} + \frac{1}{12} \sum_{i=1}^3 \| \mc D(\sigma_i) \|_1\;.
\label{eq:optfidUB0}
\end{equation}
Since by definition $\upsilon_i = \| \mc D(\sigma_i) \|_1/2$, the recovery operation in Eq.~\eqref{eq:qubit:recovery} saturates the upper bound and therefore is optimal. In the case discussed in the text, $\upsilon_i = \lambda^2$, as it can be explicitly verified: 
\begin{equation}
 \frac i2 \effdec (\hat m_2 \hat m_3+ \hat m_1 \hat m_4) = \lambda^2 \frac i2 (\hat m_2 \hat m_3+ \hat m_1 \hat m_4),
\end{equation}
(similar results hold for all qubit operators).

\section{Microscopic derivation of a Markovian master equation for a Kitaev chain in bosonic environment \label{app:mastereq2}}

In this appendix we present a weak-coupling derivation of a Markovian master equation for the system and environment setup discussed in Section~\ref{sec:bosonic}.
The derivation follows the general approach presented in \citep{petruccione} for spin systems. The discussion applies unchanged to our fermionic system, as all relevant operators are bosonic (i.e. even-degree monomials of Fermi fields and linear combinations thereof), and thus effectively behave like spins for our purposes.
Similar results have been recently presented for master equations with Lindblad operators linear in the fermionic fields in Ref.~\cite{Campbell_2015} (see also Ref.~\cite{tesi:ippoliti}).

\subsection{Generalities}

We start from an uncorrelated system-environment state $\hat \rho(0)$, and assume that the environment is in a thermal state at a generic inverse temperature $\beta$, i.e. $\hat \rho(0) = \hat \rho_S (0) \hat \rho_E(0)$ with $\rho_E(0) = \rho_E^{\text{th}}(\beta) \doteqdot e^{-\beta \mc H_{\text{env}} } / \tr \big[ e^{-\beta \mc H_{\text{env}} }\big] $. The evolution equation in the interaction picture is 
\begin{equation}
\frac{\rm d}{{\rm d}t} \hat \rho(t) = - i [\hat H_I(t), \hat \rho(t)]\;;
\end{equation}
integrating it for $\rho(t)$ and plugging the result again into the right-hand side yields
\begin{equation}
\frac{\rm d}{{\rm d}t} \hat \rho(t) = - i [\hat H_I(t), \hat \rho(0)] - \int_0^t {\rm d}s\ \left[ \hat H_I(t) , [\hat H_I(s), \hat \rho(s) ] \right] \;.
\label{eq:app-meqs2}
\end{equation}
The interaction Hamiltonian has the following structure: ${\hat H}_I(0) = \sum_\alpha \hat A_\alpha \hat B_\alpha$, where the $\{ \hat A_\alpha\}$ are system operators and the $\{\hat B_\alpha\}$ are environment operators, such that $\hat A_\alpha \hat B_\alpha$ is bosonic $\forall \alpha$ (i.e. for a given $\alpha$, $\hat A_\alpha$ and $\hat B_\alpha$ are either both bosonic or both fermionic).
This represents no loss of generality, since the fermionic parity superselection rule forces all observables to be bosonic operators, and $\hat H_I$ is an observable.
Since the Hamiltonian must be Hermitian, one can always choose $\hat B_\alpha$ to be Hermitian, while $\hat A_\alpha$ is either Hermitian (bosonic case) or anti-Hermitian (fermionic case), so that $(\hat A_\alpha \hat B_\alpha)^\dagger = (\hat B_\alpha) (\pm \hat A_\alpha) = \hat A_\alpha \hat B_\alpha$.
We remark that both the $\hat A_\alpha$ and the $\hat B_\alpha$ can be taken traceless, up to a redefinition of the system and environment Hamiltonians.
This implies that, when the environment is traced out in Eq.~\eqref{eq:app-meqs2}, the first term vanishes.
Then tracing out the environment and invoking the Born approximation to replace $\hat \rho(s)$ by the instantaneous value $\hat \rho(t)$ yields
\begin{equation}
\frac{\rm d}{{\rm d}t} \hat \rho(t) = - \int_0^\infty {\rm d}s\ \tr_E \left[ \hat H_I(t) , [\hat H_I(t-s), \hat \rho(t) ] \right] \;.
\label{eq:app-meqs3}
\end{equation}
We also removed the dependence on the initial time by integrating over the entire past history $s\to \infty$;
it is part of the assumptions of Markovianity that this does not change the result significantly.
Introducing a Fourier decomposition of the $\{\hat A_\alpha\}$ system operators such that $\hat A_\alpha(t)  = \sum_\omega \hat A_\alpha(\omega) e^{-i\omega t}$ allows us to write
\begin{align}
\frac{\rm d}{{\rm d}t} \hat \rho(t) 
& = - \sum_{\alpha,\omega} \sum_{\alpha', \omega'} \int_0^\infty {\rm d}s\ e^{-i\omega t} e^{-i\omega'(t-s)}  \nn\\
& \quad \tr_E \left[ \hat A_\alpha(\omega) \hat B_\alpha(t) , [\hat A_{\alpha'} (\omega') \hat B_{\alpha'}(t-s), \hat \rho(t) ] \right] \;.
\label{eq:app-meqs4}
\end{align}
At this point we invoke the {\it rotating wave approximation} to argue that all terms with $\omega+\omega'\neq 0$ cancel.
Introducing the environment correlation function 
\begin{equation}
\Gamma_{\alpha \beta} (\omega) 
\doteqdot \int_0^\infty {\rm d}s\ e^{i\omega s} \frac{ \tr_E \big[ \hat B_\alpha(t) \hat B_\beta (t-s)e^{-\beta \mc H_{\text{env}}} \big]}{\tr_E \big[ e^{-\beta \mc H_{\text{env}}} \big]}
\end{equation}
(which does not depend on $t$ because all involved Hamiltonians are time-independent), we have
\begin{align}
\frac{\rm d}{{\rm d}t} \hat \rho
& = - \sum_{\alpha,\beta, \omega} \Gamma_{\alpha \beta}(\omega) 
\Big( \hat A_\alpha(\omega) \hat A_\beta^\dagger(\omega) \hat \rho
- \hat A_\beta^\dagger(\omega) \hat \rho  \hat A_\alpha(\omega) 
\nn\\
& \qquad 
-  \hat A_\alpha(\omega)  \hat \rho \hat A_\beta^\dagger(\omega) 
+  \hat \rho \hat A_\beta^\dagger(\omega)\hat A_\alpha(\omega)   \Big) \;.
\label{eq:app-meqs5}
\end{align}

By finally introducing $\gamma_{\alpha\beta}(\omega) \doteqdot \Gamma_{\alpha \beta}(\omega) +\Gamma_{\beta\alpha}(-\omega)$ and $S_{\alpha \beta}(\omega) \doteqdot \frac{1}{2i} (\Gamma_{\alpha \beta}(\omega) - \Gamma_{\beta\alpha}(-\omega))$,
we can recast \eqref{eq:app-meqs5} in the form
\begin{align}
\frac{\rm d}{{\rm d}t} \hat \rho
& = - \sum_{\alpha,\beta, \omega} \gamma_{\alpha \beta}(\omega) 
\Big( \hat A_\alpha(\omega) \hat \rho  \hat A_\beta^\dagger(\omega) 
- \frac{1}{2} \left\{ \hat A_\beta^\dagger(\omega)\hat A_\alpha(\omega) , \hat \rho \right\} \Big)
\nn\\
& \qquad 
+ i S_{\alpha \beta} (\omega) \left[ \hat A_\alpha(\omega) \hat A_\beta^\dagger(\omega) , \hat \rho \right]
\label{eq:app-meqs6}
\end{align}
Here
$ \sum_{\alpha,\beta,\omega} S_{\alpha \beta} (\omega) \hat A_\alpha(\omega) \hat A_\beta^\dagger(\omega) \doteqdot \hat H_{LS}$ is the ``Lamb shift'' Hamiltonian, which renormalizes the system Hamiltonian.
As for the other terms, they contribute to the dissipative part of the dynamics
the Lindblad operators are obtained upon diagonalizing the Hermitian matrices $\gamma_{\alpha \beta}(\omega)$:
\begin{equation}
\hat L_{\alpha,\omega} = \sum_\beta \sqrt{d_{\alpha\alpha}(\omega) }c_{\alpha \beta} (\omega) \hat A_\beta(\omega)\;,
\label{eq:app-meqlind}
\end{equation}
where for each $\omega$, $c(\omega)$ is the unitary which diagonalizes $\gamma(\omega)$ into $d(\omega)$: $\gamma(\omega) = c(\omega)^\dagger d(\omega) c(\omega)$.
$\omega$ ranges over all the possible gaps in the system Hamiltonian, and the Fourier component $\hat A_\alpha(\omega)$ is constructed as 
\begin{equation}
\hat A_\alpha(\omega)
= \sum_{E_1, \, E_2:\, E_1-E_2=\omega} \hat \Pi(E_1) \hat A_\alpha \hat \Pi(E_2)\;,
\end{equation}
$E_1$, $E_2$ being energy levels for the system Hamiltonian and $\hat \Pi(E)$ being the projector on the $E$-energy eigenspace.

\subsection{Derivation of the master equation studied in Sec.~\ref{sec:bosonic}}

The decomposition of the interaction Hamiltonian \eqref{eq:bose-hint} is as follows:
$\hat{\mc H}_{\text{int}} = \sum_i \hat A_i \hat B_i$, with $\hat A_i = 2\hat a_i^\dagger \hat a_i -\id$ and $\hat B_i = -\eta \sum_n \hat X_{i,n}$.
The time evolution induced by the environment Hamiltonian $\hat{\mc H}_{\text{env}} = \sum_{i,n} \frac{1}{2} \omega_{i,n} \hat Z_{i,n}$ on the $\hat B_i$ operators yields
\begin{align}
\hat B_i(\tau) 
& = -\eta \sum_{n=1}^M e^{-i \frac{1}{2} \omega_{i,n} \tau \hat Z_{i,n}} \hat X_{i,n} e^{i \frac{1}{2} \omega_{i,n} \tau \hat Z_{i,n}} \nn \\
& = -\eta \sum_{n=1}^M \left( \cos(\omega_{i,n} \tau) \hat X_{i,n}+ \sin(\omega_{i,n} \tau) \hat Y_{i,n}  \right)\;.
\end{align}
With this information, and by factoring the thermal state as a product of one-spin operators and using the identity $e^{-\alpha \hat Z } = \cosh(\alpha) \id - \sinh(\alpha) \hat Z$, one can prove that
\begin{align}
\Gamma_{ij}(\tau)
& = \frac{ \tr \big[ \hat B_i (\tau) \hat B_j(0) e^{-\beta \hat {\mc H}_{\text{env}}} \big] } {\tr \big[ e^{-\beta \hat {\mc H}_{\text{env}}} \big] } \nn \\
& = \eta^2 \delta_{ij} \sum_{n=1}^M [ \cos(\omega_{i,n} \tau) - i \sin( \omega_{i,n} \tau) \tanh(\beta \omega_{i,n}/2) ]\;.
\end{align}
Hence, by Fourier-transforming and averaging over the environment spectra,
\begin{align}
\overline{\Gamma}_{ij}(\omega) 
& = M \eta^2 \delta_{ij} \int_0^\infty {\rm d}\tau\ e^{i\omega \tau} \int {\rm d}E\ f(E)
\times \nonumber \\
& \qquad \times [\cos(E \tau) -i \sin(E\tau) \tanh(\beta E/2) ]
\end{align}
and finally, adding the conjugate transpose,
\begin{widetext}
\begin{align}
\label{eq:overline-gamma}
\overline{\gamma}_{ij}(\omega) 
& = M \eta^2 \delta_{ij} \int_{-\infty}^\infty {\rm d}\tau\ e^{i\omega \tau} \int {\rm d}E\ f(E) [\cos(E \tau) -i \sin(E\tau) \tanh(\beta E/2) ]= \nn \\
& = \frac{\pi}{2} M \eta^2 \delta_{ij} \int {\rm d}E\ f(E) 
\left [ \delta(\omega- E) \left(1+\tanh \left(\frac{\beta E}{2} \right) \right) +
 \delta(\omega+E) \left(1-\tanh \left(\frac{\beta E}{2} \right ) \right) \right]\;;
\end{align}
\end{widetext}
as $f(E)$ is a density of state supported in $E>0$, only one of the two delta functions will contribute, depending on the sign of $\omega$, therefore the result is
\begin{align}\label{eq:gammaMR}
\overline{\gamma}_{ij}(\omega) 
& = g^2 \delta_{ij} 
\begin{cases}
f(|\omega|) \frac{1+\tanh(\beta |\omega|/2)}{2} & \text{ if } \omega > 0\\ 
f(0) & \text{ if } \omega = 0 \\
f(|\omega|) \frac{1-\tanh(\beta |\omega|)/2)}{2} & \text{ if } \omega < 0
\end{cases}
\end{align}
where $g^2 \doteqdot \pi M \eta^2$ is the effective coupling constant, which is assumed to stay finite as $M \to \infty$ and $\eta \to 0$.
This matrix is already diagonal in the $\hat A_i$ operators, so that the resulting set of Lindblad operators is given by $\hat L_{i,0} = g \sqrt{f(0)} \hat A_i(0) $ and (for $\omega \neq 0$) $\hat L_{i,\pm|\omega|} = g \sqrt{f(|\omega|) \frac{1\pm \tanh(\beta|\omega|)}{2} }  \hat A_i(\pm |\omega| )$.
For notational ease, let us introduce the function $\phi(\omega)$ via $\overline{\gamma}_{ij} (\omega) \doteqdot \phi(\omega) \delta_{ij}$, i.e.
\begin{equation}
\phi(\omega) \doteqdot g^2 f(|\omega|) \frac{1+\tanh(\beta\omega/2)}{2}\;.
\end{equation}

It is easy to determine the Fourier decomposition of the $\hat A_i$ operators: one just needs to write them in terms of $\hat b^{(\dagger)}$ operators, and count the energy quanta introduced into the system. Namely,
\begin{equation}
\left\{
\begin{aligned}
\hat A_1 & = i \hat m_1 \hat \gamma_{1,1} = i \hat m_1 (\hat b_1 + \hat b_1^\dagger)\;, \\
\hat A_{1<i<L} & = i \hat \gamma_{i-1,2} \hat \gamma_{i,1} =  (\hat b_{i-1} - \hat b_{i-1}^\dagger ) (\hat b_i + \hat b_i^\dagger ) \;,\\
\hat A_L & = \hat m_2 \hat \gamma_{L-1,2} = i \hat m_2 (\hat b_{L-1} - \hat b_{L-1}^\dagger)\;.
\end{aligned}
\right.
\end{equation}
This leads to the following list of Lindblad operators:
\begin{equation}
\left\{
\begin{aligned}
L_i^{(0)} & = \sqrt{2\phi(0)} \left( \hat b_i \hat b_{i-1}^\dagger + \hat b_{i-1} \hat b_i^\dagger  \right) \\
L_1^{(+\Delta)} & = \sqrt{\phi(\Delta)} \, i \hat m_1 \hat b_1 \\
L_1^{(-\Delta)} & =  \sqrt{\phi(-\Delta)} \, i \hat m_1 \hat b_1^\dagger \\
L_{L}^{(+\Delta)} & = \sqrt{\phi(\Delta)} \, i \hat m_2 \hat b_{L-1}\\
L_L^{(-\Delta)} & = \sqrt{\phi(-\Delta)}\,  i \hat m_2 \hat b_{L-1}^\dagger \\
L_i^{(+2\Delta)} & =  \sqrt{\phi(2\Delta)}\,  \hat b_{i-1} \hat b_i \\
L_i^{(-2\Delta)} & = \sqrt{ \phi(-2\Delta)}\, \hat b_i^\dagger \hat b_{i-1}^\dagger 
\end{aligned}
\right.
\end{equation}
where $i$ ranges from 2 to $L-1$.

As for the correction to the Hamiltonian,
it can be seen by a computation analogous to the one leading to \eqref{eq:overline-gamma} that the $S$ term in Eq.~\eqref{eq:app-meqs6} is
\begin{equation}
\overline{S}_{ij}(\omega) 
= \frac{g^2}{\pi} \delta_{ij} \mc P \int_0^\infty \frac{{\rm d}E}{|\omega|-E} f(E) \frac{\omega+E \tanh(\beta E/2)}{|\omega|+E}
\label{eq:integrals:for:LS}
\end{equation}
with $\mc P$ denoting the Cauchy principal part of the integral.
Therefore, letting $\overline{S}_{ij}(\omega) \doteqdot S(\omega) \delta_{ij}$, we have, up to additive constants,
\begin{align}
\hat{\mc H}_{LS} = 
& \ (S(\Delta)-S(-\Delta)-S(0) + S(2\Delta) )  ( \hat b_1^\dagger \hat b_1 + \hat b_{L-1}^\dagger \hat b_{L-1} )  \nonumber \\
& + 2(S(2\Delta)-S(0))\sum_{i=2}^{L-2} \hat b_i^\dagger \hat b_i  \nonumber \\
& +(2S(0) - S(2\Delta) - S(-2\Delta)) \sum_{i=2}^{L-1} \hat b_{i-1}^\dagger \hat b_{i-1} \hat b_i^\dagger \hat b_i \;.
\label{eq:LShamiltonian}
\end{align}
This renormalizes the energy gap by an amount that differs between the edge bonds and the interior bonds, and adds quartic interactions between nearest-neighbor bonds.
The picture simplifies considerably when one takes $\beta = 0 $ (i.e. $T=\infty$), since in that case $S(\omega) + S(-\omega) = 0 \ \forall \, \omega$;
thus no interactions emerge and one only has a renormalization of the Hamiltonian gaps:
\begin{align}
\hat{\mc H}_{LS} = 
& \ (2S(\Delta) - S(2\Delta))  ( \hat b_1^\dagger \hat b_1 + \hat b_{L-1}^\dagger \hat b_{L-1} )  \nonumber \\
& + 2S(2\Delta) \hat{\mc H}_K \;.
\label{eq:infiniteT-LShamiltonian}
\end{align}

\section{Parity-preserving noise models \label{app:pp}}

In this Appendix we prove a lemma that characterizes all parity-preserving Markovian noise models. 
Parity preservation (PP) for a generic channel $\mc D_t$ is stated as follows:
\begin{equation}
\average{\hat P_f}_t  = \average{\hat P_f}_0 \quad \forall \, t\; ,
\end{equation}
which is equivalent to
\begin{equation}
	\partial_t \average{\hat P_f}_t = 
	\tr \big[\hat P_f \partial_t \hat \rho \big] = 
	\tr \big[\hat P_f \mc L(\hat \rho) \big] = 
	\tr \big[\hat \rho \mc L^*(\hat P_f) \big] = 0 \;,
	\label{eq:app-pp1}
	\end{equation}
for any state $\hat \rho$.
We shall prove that a Markovian noise model is PP if and only if its associated Lindblad operators are all bosonic (BL).

{(PP) $\implies$ (BL):}
A state $\hat \rho$ is a bosonic operator and as such it can be written as $\hat \rho + \hat P_f \hat \rho \hat P_f$. Thus PP implies that 
$\tr \big[\hat \rho (\mc L^* (\hat P_f) +\hat P_f \mc L^* (\hat P_f) \hat P_f ) \big] = 0$ for any state $\hat \rho$, i.e.
	\begin{equation}
	\hat P_f \mc L^*(\hat P_f) + \mc L^*(\hat P_f) \hat P_f = 0\;.
	\end{equation}
	Decomposing the Lindblad operators in their fermionic and bosonic parts, $\hat L_i = \hat L_i^f + \hat L_i^b$, and exploiting the fact that $[\hat L_i^b,\hat P_f] = 0$ while $\{\hat L_i^f,\hat P_f\} = 0$, one gets
	\begin{align}
	\sum_i \Big( & (\hat L_i^b - \hat L_i^f)^\dagger (\hat L_i^b + \hat L_i^f) + (\hat L_i^b + \hat L_i^f)^\dagger (\hat L_i^b - \hat L_i^f)  \nn \\
	& - \hat L_i^\dagger \hat L_i  - (\hat L_i^b - \hat L_i^f)^\dagger (\hat L_i^b - \hat L_i^f)  \Big)  = 0\;,
	\end{align}
	whose trace yields $\sum_i \tr [ (\hat L_i^f)^\dagger \hat L_i^f ] =0$.
	Now, since $\tr (\hat O^\dagger \hat O)  = \| \hat O\|^2_{HS}$ is the squared Hilbert-Schmidt norm of operator $\hat O$, $\sum_i \| \hat L_i^f \|^2_{HS} = 0$ implies $\hat L_i^f = 0$ $\forall \, i$, i.e. all Lindblad operators are purely bosonic (BL).

{(BL) $\implies$ (PP):}
	From \eqref{eq:app-pp1} we have that the noise is (PP) if $\mc L^* (\hat P_f) = 0$. But using (BL) to commute $\hat P_f$ with all Lindblad operators, we have that $\mc L^* (\hat P_f) = \hat P_f \mc L^*(\id)$, and $\mc L^*(\id) = \sum_i (\hat L_i^\dagger \hat L_i -\frac{1}{2} \{ \id, \hat L_i^\dagger \hat L_i \}) = 0$; thus average parity is constant (PP).

\section{Best recovery operation of two decoupled Kitaev chains in presence of parity-preserving noise}\label{app:best:best:best}

In this Appendix we prove a generalization of a result presented in Ref.~\cite{Mazza:2013}. We consider a system composed of two Kitaev wires as discussed in Sec.~\ref{sec:ex:qq} and~\ref{sec:bosonic}; the noise model $\mathcal D_t$ is parity-preserving and does not couple the two wires. Under these assumptions it is possible to characterize the best recovery operation and its fidelity.

Let us start from the general relation:
\begin{equation}
 F(\mathcal R_t) \leq \frac 12 + \frac 1{12} \sum_{\alpha = x,y,z} \| \mathcal D_t (\hat \sigma_\alpha^L) \|_{1}
 \label{eq:optfidUB}
\end{equation}
The decoherence channel acts on two Kitaev chains, which are separated and decoupled; the noise does not couple them and conserves the parity of the fermionic number on each chain.
Decoupling the chains has tremendous importance:
for all times $t>0$, the following holds:
\begin{equation}
 \|\mathcal D_t (\hat \sigma_z^L) \|_{1} = 
 \|\mathcal D_t ( \ket{0_L} \hspace{-0.1cm} \bra{0_L}) - \mathcal D_t (\ket{1_L} \hspace{-0.1cm} \bra{1_L}) \|_{1} =
 2\;.
 \label{eq:sigma3preserved}
 \end{equation}
 
 \textit{Proof:}
 As the decoherence channel is parity preserving and does not couple the chains, $\mathcal D_t ( \ket{0_L} \hspace{-0.1cm} \bra{0_L}) $ and $\mathcal D_t (\ket{1_L} \hspace{-0.1cm} \bra{1_L}) $ have an even (odd, respectively) number of particles in each chain. 
 Therefore they are supported in orthogonal subspaces, and the trace norm is thus additive:
 \begin{equation}
 \|\mathcal D_t (\hat \sigma_z^L) \|_1= 
\|\mathcal D_t ( \ket{0_L} \hspace{-0.1cm} \bra{0_L}) \|_1 + \|\mathcal D_t ( \ket{1_L} \hspace{-0.1cm} \bra{1_L}) \|_1\;.
 \end{equation}
 Now, $\mathcal D_t ( \ket{0_L} \hspace{-0.1cm} \bra{0_L}) $ is a quantum state, and in particular a positive matrix; 
 so its trace norm is simply its trace. As $\mathcal D_t$ is trace-preserving, one has $\|\mathcal D_t ( \ket{0_L} \hspace{-0.1cm} \bra{0_L}) \|_1 = 1$. The same holds for $\mathcal D_t ( \ket{1_L} \hspace{-0.1cm} \bra{1_L})$, which concludes the proof of \eqref{eq:sigma3preserved}.
 $\blacksquare$

Now let us determine the form of the optimal recovery operation.
In the process, we shall also determine the value of $\| \mc D_t (\lpo{1}) \|_1  $ and $\| \mc D_t (\lpo{2}) \|_1 $.

In the notation introduced in \eqref{eq:recovery}, one needs to determine the optimal $\{ \hat H_i \}$ matrices.
A set of recovery matrices that saturates the upper bound \eqref{eq:optfidUB} is given by $\hat H_i = \text{sign} (\mc D_t(\lpo{i}) )$, where the sign of a Hermitian matrix is defined as $\text{sign}(A) = \sqrt{A^2} \cdot A^{-1}$.
Unfortunately, this set is not guaranteed to induce a physical map, i.e. the upper bound \eqref{eq:optfidUB} might be unattainable.
However, we shall argue, following \citep{Mazza:2013}, that in our case the matrices $\{ \hat H_i\}$ obey the Pauli matrix algebra $\hat H_j \hat H_k = i \epsilon_{jkl} \hat H_l$, and this guarantees that the induced recovery map is physical.

For the operator $\lpo{3}$, we have
\begin{equation}
  \hat H_z = \text{sign} (\mc D_t(\lpo{3}) ) = \hat \Pi_a^+ \hat \Pi_b^+ - \hat \Pi_a^- \hat \Pi_b^-\;,
  \label{eq:hz}
\end{equation}
where $\hat \Pi_a^\pm$ is the projector on the $\pm$ fermionic number parity sector of the first chain, dubbed $a$, and $\hat \Pi_b^\pm$ is the same operator for the second chain, dubbed $b$.
It is easy to verify that $\text{tr}\big[\hat H_z \mc D_t(\lpo{3})  \big] = \| \mc D_t(\lpo{3})  \|_1$.

Let us now consider the action of the decoherence channel on the operator $\hat \sigma^L_1$, which is best analyzed through the the operators $\hat R_{a}$ and $\hat R_b$:
\begin{equation}
\hat R_a = \mc D_t(\ketbra{0_{a}}{1_a} )\;, \quad 
\hat R_b = \mc D_t(\ketbra{0_b}{1_{b}} )\;. 
\label{eq:Rdef}
\end{equation}
where $\ket{0_j}$ and $\ket{1_j}$ denote states of the chain $j$ such that: $\ket{0_L} = \ket{0_a} \otimes \ket{0_b}$.
In terms of these operators, one has
\begin{equation}
 \mathcal D_t (\lpo{1} ) = 
 \mathcal D_t ( \ketbra{0_L}{1_L} + \ketbra{1_L}{0_L} )  =  \hat R_a \hat R_b + \hat R_b^\dagger \hat R_a^\dagger\;.
 \label{eq:Dx:str}
\end{equation}
Let us take the square of this operator:
\begin{align}
\left(\mc D_t( \hat \sigma_1^L) \right)^2 
 =& \left(\hat R_a \hat R_b + \hat R_b^\dagger \hat R_a^\dagger \right)^2 
= - \hat R_a^2 \hat R_b^2 -  \hat R_b^{\dagger2}\hat R_a^{\dagger2} +\nn \\
&+ \hat R_a \hat R_a^\dagger \hat R_b \hat R_b^\dagger +
  \hat R_a^\dagger \hat R_a \hat R_b^\dagger \hat R_b\;;
\label{eq:DxSquared1}
\end{align}
This holds because $\hat R_a$ and $\hat R_b$ are fermionic operators with disjoint supports, and thus anti-commute. 
Since by definition $\hat R_a = \hat \Pi_a^+ \hat R_a \hat \Pi_a^- $,
it squares to zero $\hat R_a^2 = \hat \Pi_a^+ \hat R_a \hat \Pi_a^-  \hat \Pi_a^+ \hat R_a \hat \Pi_a^-  = 0$; thus
\eqref{eq:DxSquared1} simplifies to
\begin{equation}
(\mc D_t( \hat \sigma_1^L) )^2 = \hat R_a \hat R_a^\dagger \hat R_b \hat R_b^\dagger + \hat R_a^\dagger \hat R_a \hat R_b^\dagger \hat R_b\;,
\label{eq:DxSquared2}
\end{equation}
where we note that the first and second terms are positive Hermitian operators on the subspaces $\hat \Pi_a^+ \hat \Pi_b^+ $ and $\hat \Pi_a^- \hat \Pi_b^- $ respectively.
Taking the operator square root now yields
\begin{equation}
| \mc D_t(\hat \sigma_1^L) | = \sqrt{ \hat R_a \hat R_a^\dagger}  \sqrt{ \hat R_b \hat R_b^\dagger} + \sqrt{ \hat R_a^\dagger \hat R_a}  \sqrt{ \hat R_b^\dagger \hat R_b} \;,
\end{equation}
and the operator $\sqrt{ \hat R_a \hat R_a^\dagger}$ coincides the positive Hermitian operator $\hat P_a$ defined by the polar decomposition $\hat R_a = \hat U_a \hat P_a$ (the same holds for chain $b$).
Finally, we can compute the trace norm via the definition $\tracenorm{A} = \tr{|A|}$:
\begin{equation}
 \| \mc D_t(\hat \sigma_1^L) \|_1 = 2 \text{tr}_a (\hat P_a) \text{tr}_b ( \hat P_b)\;,
 \end{equation}

Now we have that $| \mc D_t(\lpo{1}) | = | \mc D_t (\lpo{2}) |$ , as can be seen by plugging $\lpo{2} = -i\ketbra{0_a0_b}{1_a1_b} +h.c.$ instead of $\lpo{1} = \ketbra{0_a0_b}{1_a1_b} + \mathrm{H.c.}$ into equation \eqref{eq:DxSquared2}.
Therefore $\tracenorm{\mc D_t(\lpo{1})} = \tracenorm{\mc D_t (\lpo{2}) } $, and the optimal fidelity upper bound \eqref{eq:optfidUB} in our case where the noise has the same action on both chains reads
\begin{equation}
 F_{\rm opt} \leq \frac{2}{3} + \frac{1}{3} \left(\tr{\hat P} \right)^2\;,\
 \label{eq:optfidUBspecial}
\end{equation}
where $\hat P$ is meant as an operator on a single chain, and the trace goes over a single chain as well.

Let us now identify the optimal recovery operation, and check that the upper bound \eqref{eq:optfidUB} is indeed attainable.
The recovery matrix for $\lpo{1}$ is
\begin{align}
\hat H_1 
 = &(  \hat R_a \hat R_b + \hat R_b^\dagger \hat R_a^\dagger ) 
\cdot 
 | \hat R_a \hat R_b + \hat R_b^\dagger \hat R_a^\dagger  |^{-1}= \nonumber \\
& = \hat R_a (\hat R_a^\dagger \hat R_a)^{-1/2} \hat R_b (\hat R_b^\dagger \hat R_b)^{-1/2} +
\nn \\
& 
+ \hat R_b^\dagger (\hat R_a \hat R_a^\dagger)^{-1/2} \hat R_a^\dagger (\hat R_b \hat R_b^\dagger)^{-1/2}\;;
\end{align}
now, since $\hat R_a = \hat U_a \hat P_a$, 
we have $ \hat R_a (\hat R_a^\dagger \hat R_a)^{-1/2} = \hat \Pi_a^+ \hat U_a \hat \Pi_a^- $ and $ \hat R_a^\dagger (\hat R_a \hat R_a^\dagger)^{-1/2} = \hat \Pi_a^- \hat U_a^\dagger \hat \Pi_a^+ $ (and the same for chain $b$).
Putting it all together, we get
\begin{equation}
\hat H_1 = \hat \Pi_a^+ \hat \Pi_b^+ \hat U_a \hat U_b \hat \Pi_a^- \hat \Pi_b^- 
+  \hat \Pi_a^- \hat \Pi_b^-  \hat U_b^\dagger \hat U_a^\dagger \hat \Pi_a^+ \hat \Pi_b^+\;.
\end{equation} 
The same reasoning applied to \lpo{2} yields
\begin{equation}
\hat H_2 = -i \hat \Pi_a^+ \hat \Pi_b^+ \hat U_a \hat U_b \hat \Pi_a^- \hat \Pi_b^- 
+i  \hat \Pi_a^- \hat \Pi_b^-  \hat U_b^\dagger \hat U_a^\dagger \hat \Pi_a^+ \hat \Pi_b^+\;.
\end{equation}
Both matrices are Hermitian and by construction they have operator norm equal to 1.

At this point, recalling the form of the third recovery matrix \eqref{eq:hz}, a simple computation exploiting the properties of unitary matrices and orthogonal projectors shows that the $\{\hat H_i \}$ matrices obey the Pauli algebra, and thus by following the discussion presented in \citep{Mazza:2013} one can prove that the recovery map is CPTP, and the fidelity upper bound \eqref{eq:optfidUBspecial} is attainable.

Finally, let us recast the optimal fidelity \eqref{eq:optfidUBspecial} in a form that is more convenient for numerical computations.
Let us consider a single chain (we drop the subscript $a$ for convenience).
We have $\tr{\big[\hat P \big]} = \| \hat R \|_1$; 
now, as $\hat R$ and $\hat R^\dagger$ act on orthogonal subspaces ($ \hat R = \hat \Pi^+ \hat R \hat \Pi^-$ and $ \hat R^\dagger = \hat \Pi^- \hat R^\dagger \hat \Pi^+$), the trace norm is additive and thus
$  \| \hat R \|_1 = { \frac{1}{2} \| \hat R + \hat R^\dagger \|_1} $ .
The formula is convenient because:
\begin{equation}
\hat R + \hat R^\dagger  
= \mc D_t ( \ketbra{0}{1} + \ketbra{1}{0} ) 
= \mc D_t \big( (\hat d_0 + \hat d_0^\dagger) \hat \Pi_G \big)\;,
\end{equation} 
where $\hat d_0$ is the bi-localized Dirac zero mode
and by definition $\hat d_0 + \hat d_0^\dagger  = \hat m_1$.
Therefore, $\tr{\big[\hat P \big]} = \frac{1}{2} \| \mc D_t ( \hat m_1 \hat \Pi_G ) \|_1$, and the optimal fidelity can be rewritten as
\begin{equation}
 F^{\text{opt}}_t = \frac{2}{3} + \frac{1}{12} \left( \| \mc D_t ( \hat m_1 \hat \Pi_G ) \|_1 \right)^2\;,
\end{equation}
which proves Eq.~\eqref{eq:optfidSingleChain}.

\end{document}